\documentclass[3p,times,twocolumn]{elsarticle}
 \biboptions{comma,sort&compress}
 \usepackage[table]{xcolor}
\usepackage{graphicx}
\usepackage{here}
\usepackage{ecrc}

\volume{00}

\firstpage{1}

\journalname{Nuclear and Particle Physics Proceedings}

\runauth{}

\jid{nppp}

\jnltitlelogo{Nuclear and Particle Physics Proceedings}

\usepackage{amssymb}

\usepackage[figuresright]{rotating}

\definecolor{bluette}{rgb}{.2,.4,0}
\definecolor{salmon}{rgb}{.9,0.68,0.5}
\definecolor{motive}{rgb}{0.2,1,.5}
\definecolor{list}{rgb}{0.3,.8,.1}
\definecolor{moe}{rgb}{1,.7,.5}
\definecolor{mote}{rgb}{.7,.5,.6}
\definecolor{pisello}{rgb}{.1,1,0}
\definecolor{orange}{rgb}{1,.7,0}
\definecolor{oliva}{rgb}{.1,.5,0.3}
\definecolor{greenda}{rgb}{0,.3,.2}
\definecolor{greenli}{rgb}{0.5,.8,.0}
\definecolor{blueda}{rgb}{0,.1,.6}
\definecolor{purple}{rgb}{.7,.1,.2}
\definecolor{marrone}{rgb}{1,0.7,0}
\definecolor{pinky}{rgb}{1,0.8,0.8}
\definecolor{rose}{rgb}{1,0.4,0}

\def\beq{\begin{equation}}
\def\eeq{\end{equation}}
\def\bea{\begin{eqnarray}}
\def\eea{\end{eqnarray}}
\def\bq{\begin{quote}}
\def\eq{\end{quote}}

\def\nnb{\nonumber}
\def\ga{\left(}
\def\dr{\right)}

\def\lrar{\Longrightarrow}
\def\lrar2{\longrightarrow}
																
\def\nnb{\nonumber}

\def\la{\langle}
\def\ra{\rangle}

\def\ba{\vspace*{-0.2cm}\begin{array}}
\def\ea{\end{array}\vspace*{-0.2cm}}

\def\b{$\bullet~$}
\def\d{$\diamond~$}

\def\als{\alpha_s}

\def\gg2{\la\alpha_s G^2 \ra}
\def\gg3{g^3f_{abc}\la G^aG^bG^c \ra}
\def\ggg4{\la\als^2G^4\ra}

\def\gg{\lag g^{2}_{s} G^2 \rag}
\def\ggg{\lag g^{3}_{s}G^3\rag}

\usepackage{amsmath}
\usepackage{slashed}
\usepackage{color}

\begin{document}
\begin{frontmatter}

\title{QCD parameters and SM-high precisions from $e^+e^-\to$  Hadrons : Summary\tnoteref{invit}}
\tnotetext[invit]{Talk given at QCD23, 10-14 july 2023, Montpellier-FR}
\author{Stephan Narison
}
\address{Laboratoire
Univers et Particules de Montpellier (LUPM), CNRS-IN2P3, \\
Case 070, Place Eug\`ene
Bataillon, 34095 - Montpellier, France\\
and\\
Institute of High-Energy Physics of Madagascar (iHEPMAD)\\
University of Ankatso, Antananarivo 101, Madagascar}
\ead{snarison@yahoo.fr}


\date{\today}
\begin{abstract}
\noindent
In this talk, I summarize the results obtained recently in Ref.\,\cite{SNe} using the PDG 22 compilation of the $e^+e^-\to$  Hadrons $\oplus$ the recent CMD3 data for the pion form factor.  Using the gluon condensate $\la \alpha_s G^2\ra=(6.49\pm 0.35)\times 10^{-2}$ GeV$^4$ from heavy quark sum rules, the extracted QCD four-quark and dimension eight condensate condensates values are: $\rho\alpha_s\la\bar\psi\psi\ra^2= (5.98\pm 0.64)\times 10^{-4}$ GeV$^6$  and $d_8= (4.3\pm 3.0)\times 10^{-2}$ GeV$^8$ from the ratio ${\cal R}_{10}$ of Laplace sum rules to order $\alpha_s^4$. Inversely using these estimated values of the condensates,
we obtain from ${\cal R}_{10}$: $\la \alpha_s G^2\ra=(6.12\pm 0.61)\times 10^{-2}$ GeV$^4$ which leads to the average $(6.40\pm 0.30)\times 10^{-2}$ GeV$^4$. 
Using the lowest $\tau$-like decay moment, the mean result of  Fixed Order (FO) and Contour Improved (CI) PT series within the standard OPE is : $\alpha_s(M_\tau)=0.3385(50)(136)_{syst}$ [resp. $0.3262(37)(78)_{syst}$] to order $\alpha_s^4$ [resp. $\alpha_s^5$] leading to $\alpha_s(M_Z)$=0.1207(17)(3) [resp. 0.1193(11)(3)], while the sum of the non-perturbative contribution at $M_\tau$ is\,: $\delta^V_{NP}(M_\tau)=(2.3\pm 0.2)\times 10^{-2}$. Using the same data, one also obtains the LO hadronic vacuum polarization to the muon and $\tau$ anomalous magnetic moments: $a_\mu\vert^{hvp}_{l.o}= (7036.5\pm 38.9)\times10^{-11}, \, a_\tau\vert^{hvp}_{l.o}= (3494.8\pm 24.7)\times10^{-9} $ which leads to : $\Delta a_\mu\equiv a_\mu^{exp}-a_\mu^{th} = (143\pm 42_{th}\pm 22_{exp})\times 10^{-11}$ indicating about 3$\sigma$ discrepancy  between the SM predictions and experiment.
One also finds: $\alpha^{(5)}(M_Z)\vert_{had}=(2766.3\pm 4.5)\times 10^{-5}$. 
\begin{keyword}  QCD spectral sum rules, QCD parameters,  Hadron masses and couplings, $\tau$-decay, g-2.


\end{keyword}
\end{abstract}
\end{frontmatter}
\newpage
\section{Introduction}
\vspace*{-0.2cm}
Precise determinations of the QCD parameters and an accurate determination of the leptons anomalous magnetic moments are important inputs for testing the Standard Model (SM).  In this talk, I summarize the results obtained recently in Ref.\,\cite{SNe} (referred here and after as SN) using the PDG 22 compilation of the $e^+e^-\to$  Hadrons\,\cite{PDG} $\oplus$ the recent CMD3\,\cite{CMD3} data for the pion form factor.
\vspace*{-0.45cm}
\section{The SVZ-expansion}
\vspace*{-0.2cm}
Within the Shifman-Vainshtein-Zakharov (SVZ)\,\cite{SVZa}\,\footnote{For reviews, see e.g.\,\cite{ZAKA, SNB1,SNB2,SNREV1} and references quoted in\,\cite{SNe}.} Operator Product Expansion (OPE),  QCD condensates with higher and higher dimensions are assumed to approximate the not yet known QCD non-perturbative contributions. In the case of the two-point correlator :
\vspace*{-0.5cm}
\bea
\hspace*{-0.6cm} 
\Pi^{\mu\nu}_H(q^2)&=&i\hspace*{-0.1cm}\int \hspace*{-0.15cm}d^4x ~e^{-iqx}\la 0\vert {\cal T} {J^\mu_H}(x)\ga {J^\nu_H}_H(0)\dr^\dagger \vert 0\ra\nnb\\
&~~=-&(g^{\mu\nu}q^2-q^\mu q^\nu)\Pi_H(q^2)
 \label{eq:2-point}
 \eea
built from the T-product of the bilinear $I=1$  vector current :
\beq
 J^\mu_H(x)=\frac{1}{2}{[}: \bar\psi_u\gamma^\mu\psi_u-\bar\psi_d\gamma^\mu\psi_d:{]},
\eeq
the SVZ-expansion reads ($q^2\equiv -Q^2$)\,:
\beq
8\pi^2\Pi_H(-Q^2,m_q^2,\mu)=\sum_{D=0,2,..}\hspace*{-0.25cm}\frac{C_{D}(q^2,m_q^2,\mu)\la O_{D}(\mu)\ra}{(Q^2)^{D/2}}~, 
\label{eq:ope}
\eeq
where $m_q$ is the quark mass, $\mu$ is the subtraction scale which separates the long and short distance dynamics. $C_{D}$ are perturbatively calculable Wilson coefficients while $\la O_{D}(\mu)\ra$ are non-perturbative QCD condensates of dimension $D$. In the phenomenological analysis, the OPE is often truncated at $D=6,8$ where the approach gives a satisfactory explanation of different data (see e.g. \cite{ZAKA, SNB1,SNB2,SNREV1,SNR,BELLa}). However, one should note that the contributions of higher dimension condensates are not under a good control due to the large number of Feynman diagrams ones, to the inaccurate estimate of their size and to the difficulty to build a renormalizaion group invariant (RGI) condensate due to their mixing under renormalization\,\cite{SNTARRACH}. 

In SN, one uses the PT expression up to order $\alpha_s^4$ while the OPE is truncated at $D=6,8$. 

One should mention that, besides the well-known $\la\bar \psi\psi\ra$ quark condensate, the gluon condensates have been determined from heavy quark sum rules\,\cite{SNparam, SNcb1}:
\bea
\la\alpha_s G^2\ra =  (6.39\pm 0.35)\times 10^{-2}\,{\rm GeV^4},\nnb\\
{\la g^3  G^3\ra}/{\la\alpha_s G^2\ra}=8.2(1.0)\,[{\rm GeV^4}],
\label{eq:asg2}
\eea
while different analyzes of the light meson systems lead to the value of the four-quark condensate\,(see e.g. the papers quoted in \cite{SNREV1}) :
\beq
\rho \alpha_s\la \bar \psi\psi\ra^2=5.8(9)\times 10^{4} \,[\rm GeV^6].
\eeq
\section{QCD condensates from the ratio of LSR}
\subsection*{\b The ratio of Laplace sum rules (LSR)}
In SN, the dimension  $D=4,6$ and 8 condensates appearing in the OPE of the two-point vector correlator have been re-estimated using the ratio of Laplace sum rule moments\,\cite{SVZa,SNR,BELLa}\footnote{For a recent review, see e.g.\,\cite{SNLSR}.}:
\beq
 {\cal R}^c_{10}(\tau)\equiv\frac{{\cal L}^c_{1}} {{\cal L}^c_0}= \frac{\int_{t>}^{t_c}dt~e^{-t\tau}t\, R_{ee}^{I=1}(t,\mu) }   {\int_{t>}^{t_c}dt~e^{-t\tau} R_{ee}^{I=1}(t,\mu) },
\label{eq:lsr}
\eeq
where $\tau$ is the LSR variable, $t>$   is the hadronic threshold.  Here $t_c$ is  the threshold of the ``QCD continuum" which parametrizes, from the discontinuity of the Feynman diagrams, the spectral function  ${\rm Im}\,\Pi_H(t,m_q^2,\mu^2)$.  $m_q$ is the quark mass and $\mu$ is an arbitrary subtraction point. The spectral function is related  through the optical theorem to the isovector part of the ratio $R^{I=1}_{ee}$ as:
\beq
R^{I=1}_{ee}\equiv \frac{\sigma(e^+e^-\to\,{\rm Hadrons})}{\sigma(e^+e^-\to\mu^+\mu^-)}=\ga\frac{3}{2}\dr 8\pi\, {\rm Im} \Pi_H(t).
\eeq
\subsection*{\b Fits of the isovector $e^+e^-\to\,{\rm Hadrons}$ data}
\d Along this paper, we shall use the Mathematica program FindFit with optimized $\chi^2$ and the program NSolve for our numerical analysis. 

\d To be conservative, we shall fit separately the ensemble of high data points and low data points. Our final result will be the mean of these two extremal values. 

\begin{figure}[hbt]
\begin{center}
\includegraphics[width=7.5cm]{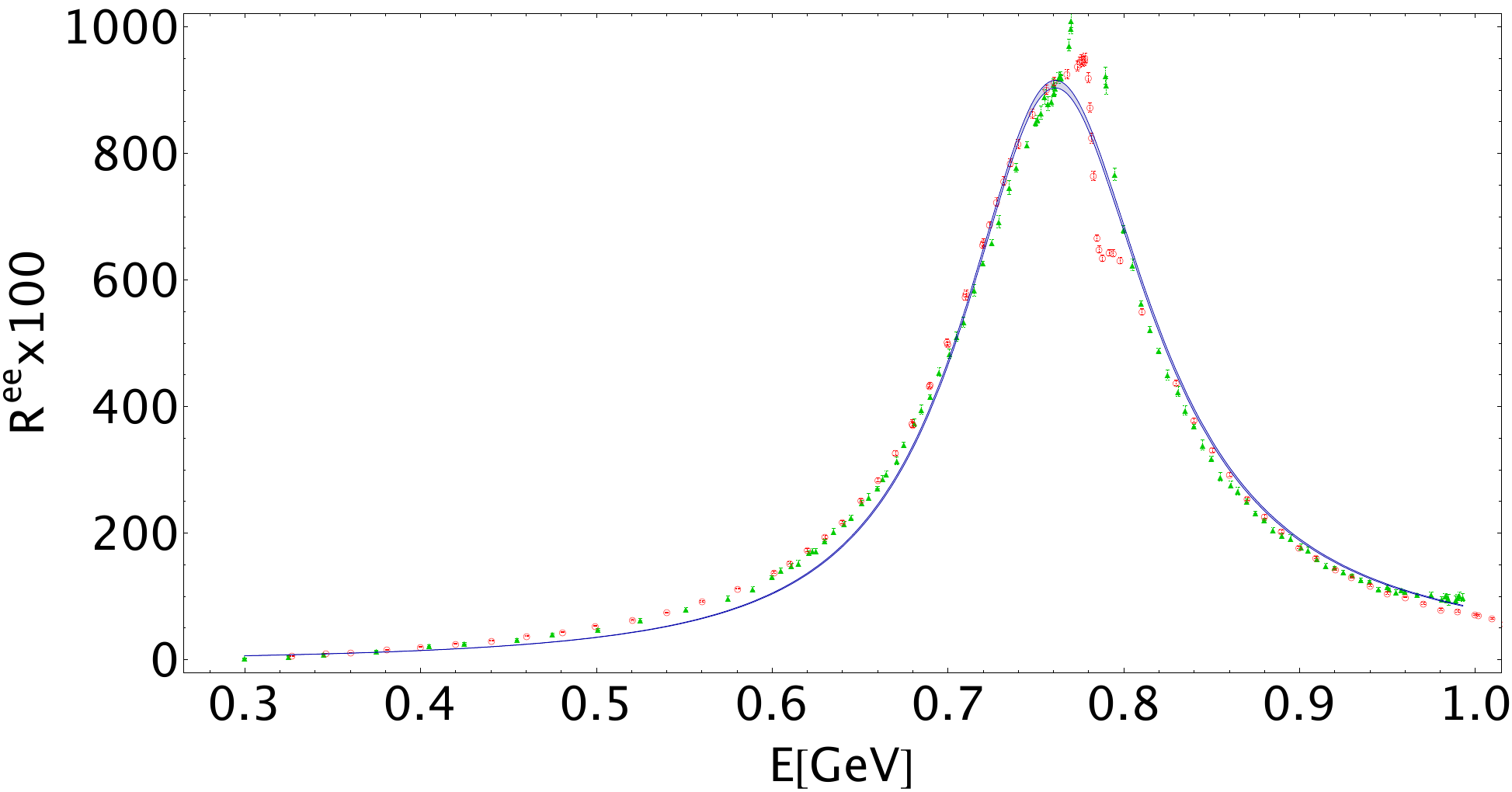}
\caption{\footnotesize  Fit of the data without $\omega$  using a minimal Breit-Wigner parametrization of the PDG\,\cite{PDG} compilation (green triangle) . A comparison with the new CMD3\,\cite{CMD3} data is given (open red circle).  } \label{fig:rho}
\end{center}
\vspace*{-0.5cm}
\end{figure}

\d From $2m_\pi$ to 0.993 GeV, we shall use a simple Breit-Wigner parametrization of the pion form factor after subtracting the $\omega(782)$ contribution using a narrow width approximation (NWA). Using as input the PDG value:
\vspace*{-0.25cm}
\beq
\Gamma(\rho\to e^+e^-)=(7.0.3\pm 0.04)\,{\rm keV},
\eeq
we deduce in units of MeV\,:
\beq
M_{\rho}=(755.605\pm 4.05)\, ,~~~~~~~~\Gamma_\rho^{\rm tot}=  (131.98\pm 0.06)\,,
\eeq
which leads to the fit in Fig.\,\ref{fig:rho}. 

\begin{figure}[hbt]
\begin{center}
\includegraphics[width=7.5cm]{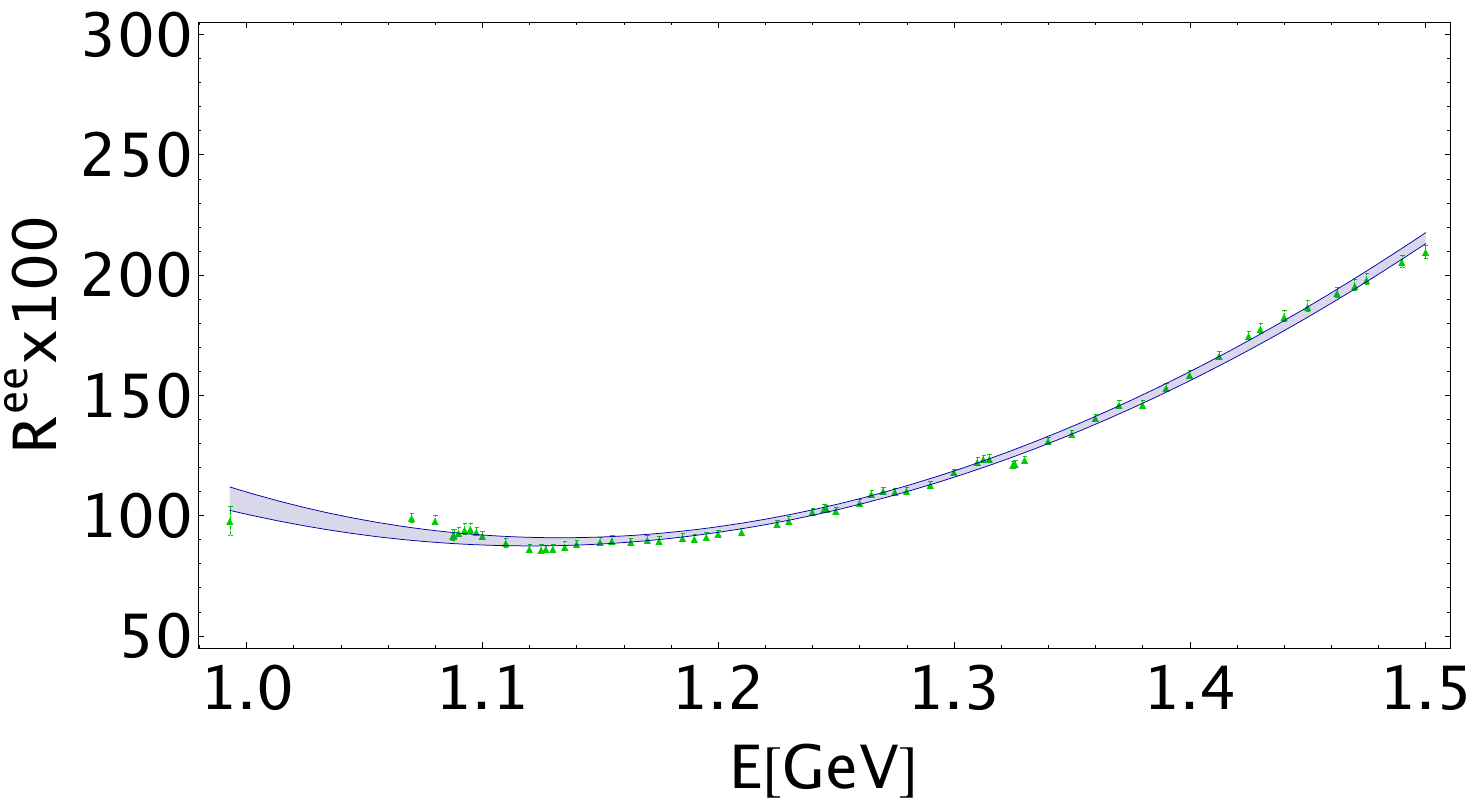}
\caption{\footnotesize  Fit of the data using a cubic polynomial  interpolation  formula.} \label{fig:3pi}
\end{center}
\vspace*{-0.5cm}
\end{figure}

\d From 0.993 to 1.55 GeV, we fit the data using a simple interpolation program with polynomials. We substract the $3\pi$ backgrounds by using the $SU(2)$ relation between the isoscalar and isovector states (a suppression 1/9 factor). 
We neglect the $\bar KK$ contributions from isoscalar sources which, in addition to the $SU(3)$ suppression factor is also suppressed by phase space. The fit is shown in Fig.\,\ref{fig:3pi}. 

\begin{figure}[hbt]
\begin{center}
\includegraphics[width=7.5cm]{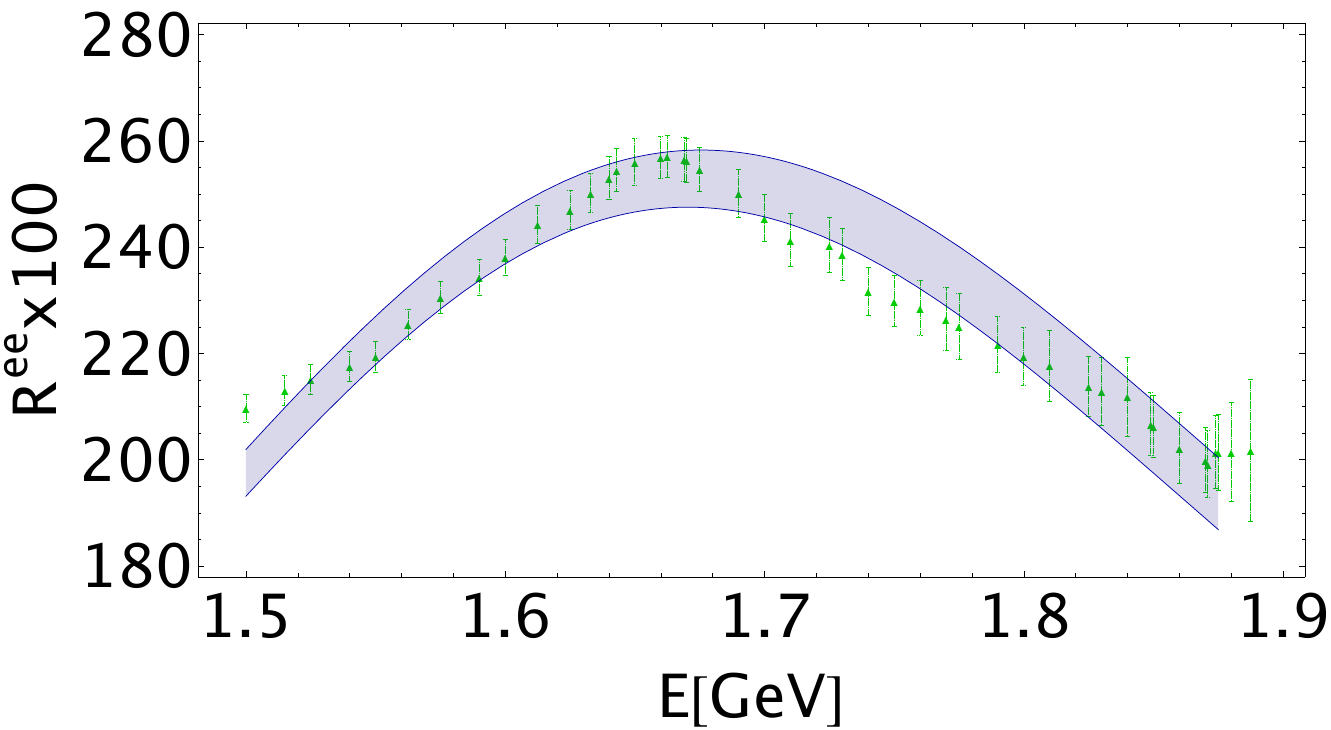}
\caption{\footnotesize  Fit of the data using a minimal Breit-Wigner parametrization.} 
\label{fig:rhoprime}.
\end{center}
\vspace*{-0.75cm}
\end{figure}

\d From 1.5 to   1.875 GeV, we use a Breit-Wigner fit for the $\rho'$ meson and obtain:
\bea
&&M_{\rho'}=1.6\,{\rm GeV}\,\nnb\\
&&\Gamma(\rho'\to e^+e^-) = [10.5\,({\rm resp.}\, 9.73)]~{\rm keV}\, .\nnb\\
&&\Gamma_{\rho'}^{\rm tot}=  [720\,({\rm resp.}\,694)]\,\rm{MeV},
\eea
from the high (resp. low) data points. The fit is shown in Fig.\,\ref{fig:rhoprime}.

\begin{figure}[hbt]
\begin{center}
\includegraphics[width=7.5cm]{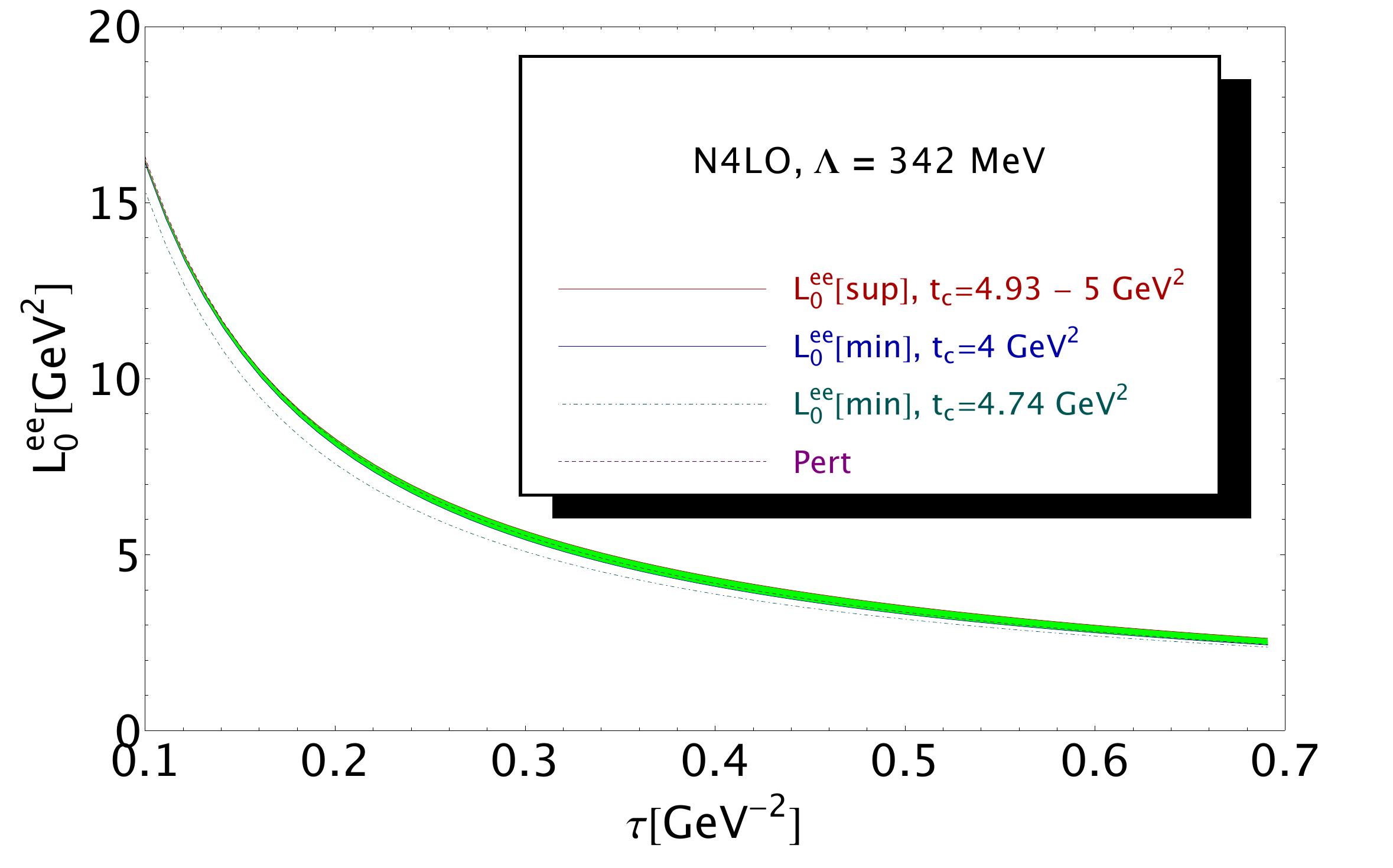}
\caption{\footnotesize  Comparison of the QCD expression of ${\cal L}^{ee}_{0}$ with the data for the obtained value of $t_c$ given in Eq.\, \ref{eq:tc}.} \label{fig:L0-tc}.
\end{center}
\vspace*{-0.75cm}
\end{figure}

\begin{figure}[hbt]
\begin{center}
\includegraphics[width=7.5cm]{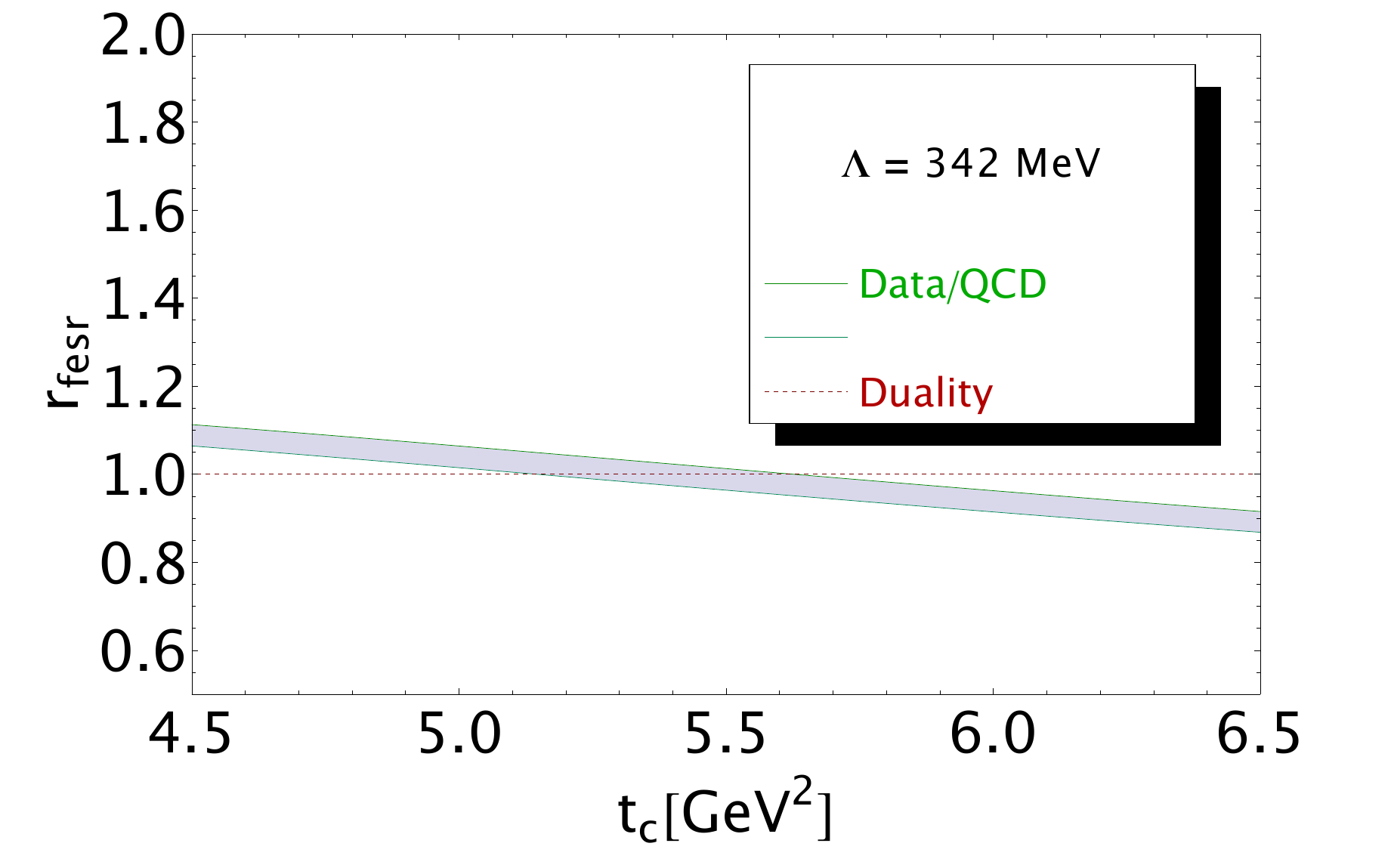}
\vspace*{-0.25cm}
\caption{\footnotesize  Variation of the ratio in Eq.\,\ref{eq:ratio-fesr} versus $t_c$.}\label{fig:fesr}
\end{center}
\vspace*{-0.75cm}
\end{figure}

\d We paramerize by the QCD continuum the data above 1.875 GeV. 
The continuum threshold $t_c$ is fixed in SN at:
\beq
t_c= (4\sim 5)~{\rm GeV}^2
\label{eq:tc}
\eeq
from the asymptotic coincidence of the experimental side and QCD side  of the lowest moment ${\cal L}_0$ from $\tau\leq 0.7$ GeV$^{-2}$ as shown in  Fig.\,\ref{fig:L0-tc}.  
This value of $t_c$ is confirmed from FESR duality constraint\,\cite{FESR} in Fig.\,\ref{fig:fesr} 
from the ratio:
\beq
r_{fesr}=\frac{\int_0^{t_c} dt\, R_{ee}^{I=1}(t)\vert_{data}} {\int_0^{t_c} dt\, R_{ee}^{I=1}(t)\vert_{qcd}},
\label{eq:ratio-fesr}
\eeq
which corresponds to $t^{fesr}_c=(5.1\sim 5.6)$ GeV$^2$ but does not favour the choice:
$
t_c=1.55~{\rm  GeV}^2,
$
used in Ref.\,\cite{BOITO} for the  unpinched moment which is  equivalent to the lowest degree of FESR used here.  

\subsection*{\b Estimate of the QCD condensates}
We attempt to extract the dimension 4,6 and 8 condensates using different strategies.
\vspace*{-0.25cm}
\subsection*{\hspace*{0.5cm}\d  $d_4$ and $d_6$ condensates}
\begin{figure}[hbt]
\begin{center}
\includegraphics[width=7.5cm]{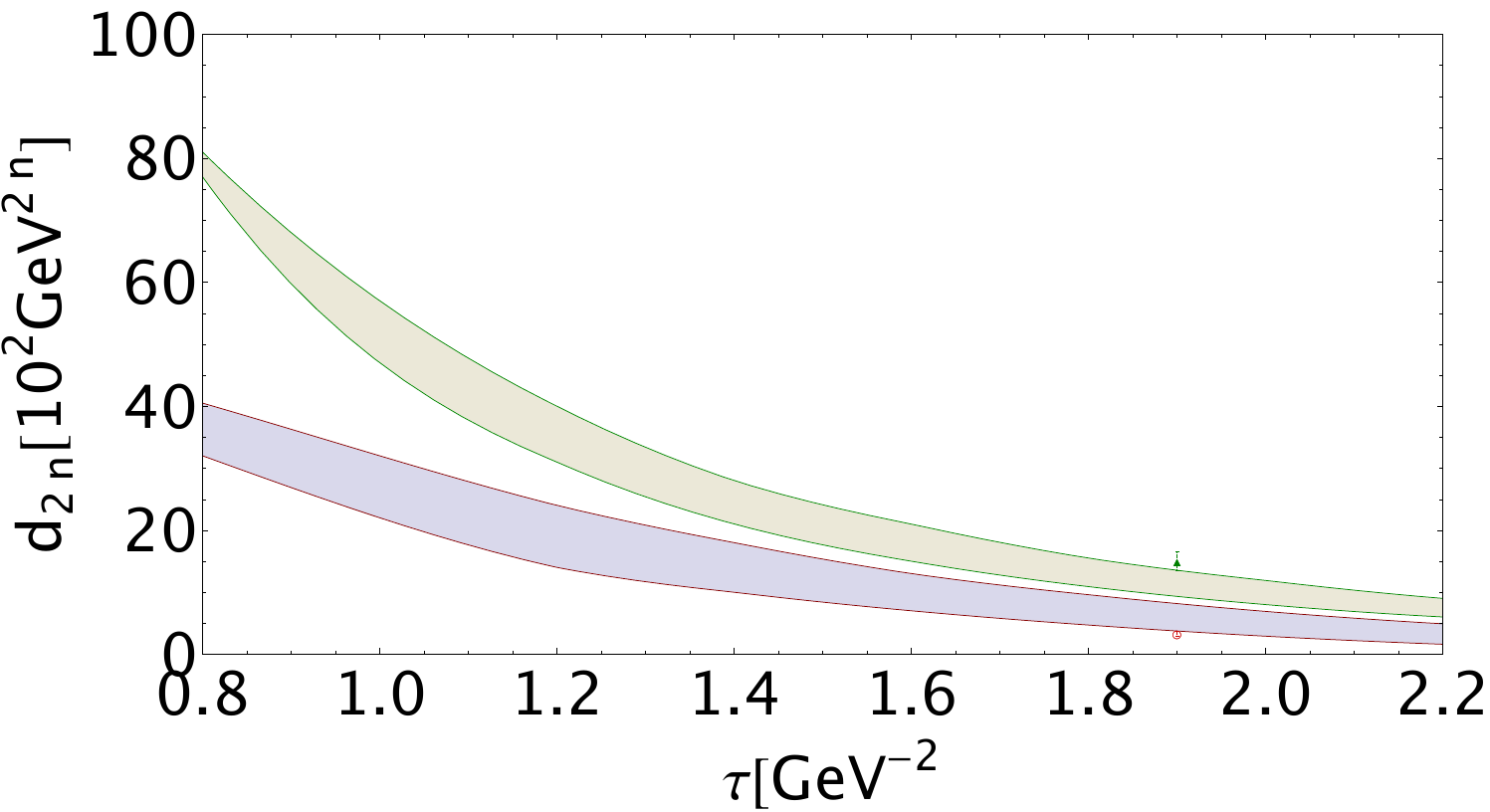}
\caption{\footnotesize  Two-parameter fit of 
$\la \alpha_sG^2\ra$ and $-d_6$ from the ratio of moments ${\cal R}_{10}$.} \label{fig:d46}.
\end{center}
\vspace*{-0.5cm}
\end{figure}

Using a two-parameter fit, we attempt to extract  in Fig.\ref{fig:d46} the $d_4$ gluon $\la\alpha_s G^2\ra$ (green curves) and $,d_6$ four-quark condensates (red curves). The analysis does not indicate $\tau$-stability but reproduces the results of Ref.\cite{LNT} at $\tau\simeq 1.9$ GeV$^{-2}$. 
\vspace*{-0.75cm}
\subsection*{\hspace*{0.5cm}\d  $d_6$ and $d_8$ condensates to order $\alpha_s^4$}
Instead, fixing the value of $\la\alpha_s G^2\ra=(6.49\pm 0.35)\times 10^{-2}$ GeV$^4$ from heavy quarkonia\,\cite{SNparam}, we estimate the $d_6$ and $d_8$ condensates to order $\alpha_s^4$. The analysis is shown in Fig\,\ref{fig:d68-as4}. 
We obtain for $\tau\simeq (2\sim 3)$ GeV$^{-2}$:
\begin{figure}[hbt]
\begin{center}
\includegraphics[width=7.5cm]{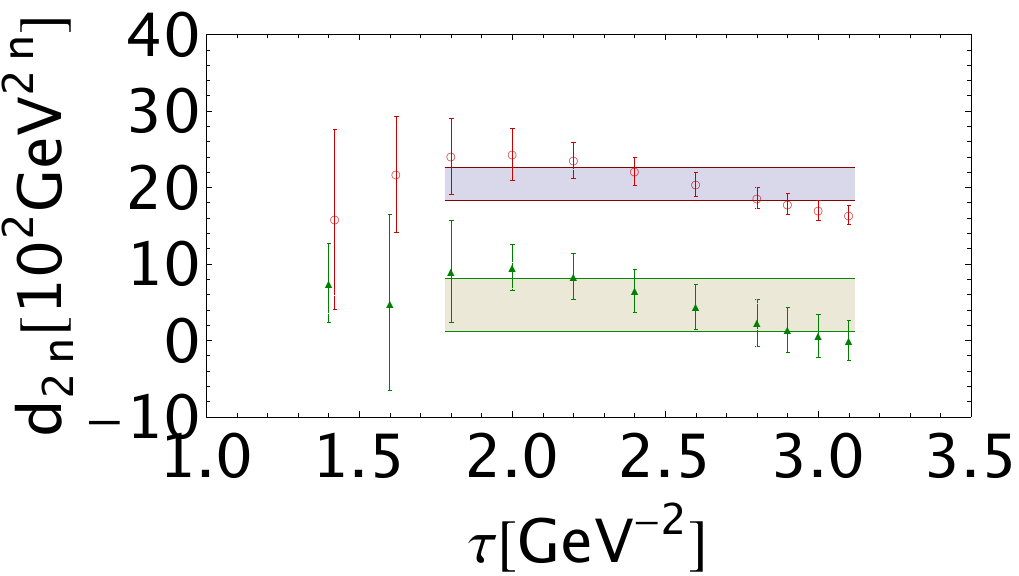}
\caption{\footnotesize  Two-parameter fit of $-d_6$ and $+d_8$ to order $\alpha_s^4$  for a given value of $\la\alpha_s G^2\ra$ from Eq.\,\ref{eq:asg2}. The red (resp. green) points are the values of $-d_6$ (resp $d_8$).} 
\label{fig:d68-as4}.
\end{center}
\vspace*{-1cm}
\end{figure} 
\bea
d_6 &=&  -(20.5\pm2.2)\times 10^{-2}\,{\rm GeV^6},\nnb\\
 d_8&=& (4.7\pm 3.5)\times 10^{-2}\,{\rm GeV^8}.
\label{eq:res-d68}
\eea
The value of $d_6$ remains unchanged from $\alpha_s^2$ to $\alpha_s^4$ which is not the case of $d_8$ which has decreased by about a factor 2.5\,! 
Then, we try to improve the determination of $d_8$ by using $d_6$ as input in addition to the previous input parameters. 
The analysis shown in Fig.\ref{fig:d8} indicates a much better stability though we do not (unfortunately) gain much on the accuracy. 
\begin{figure}[hbt]
\begin{center}
\includegraphics[width=7.5cm]{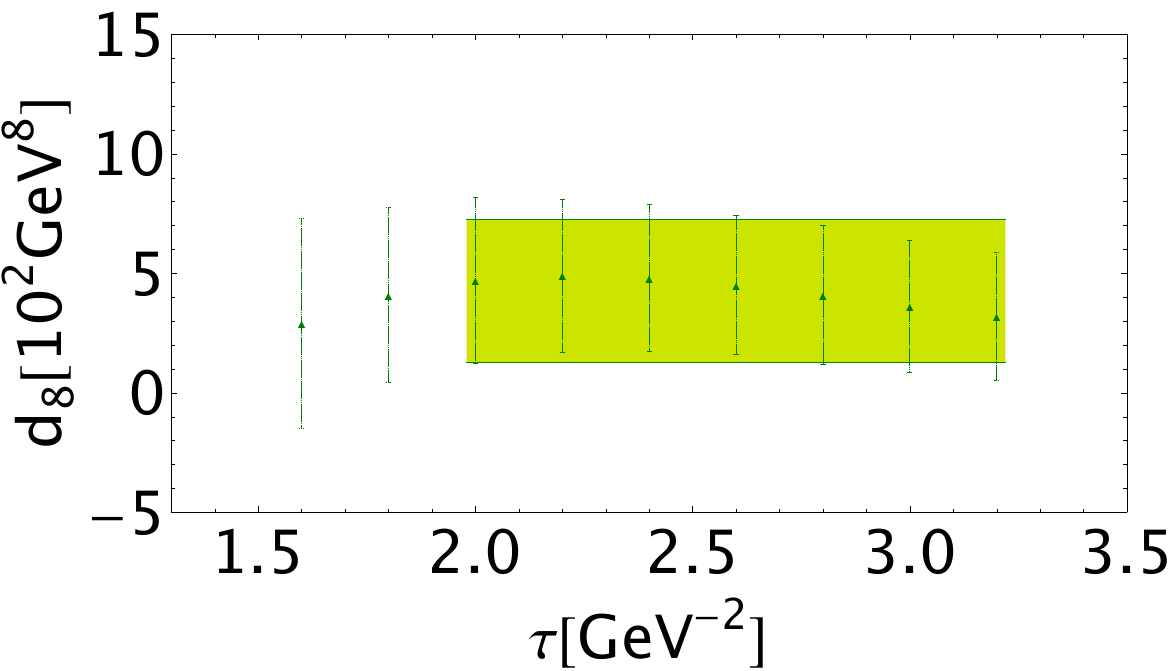}
\caption{\footnotesize  One-parameter fit of $d_8$ to order $\alpha_s^4$  for a given value of $\la\alpha_s G^2\ra$ from Eq.\,\ref{eq:asg2} and $d_6$ from Eq.\,\ref{eq:res-d68}.} 
\label{fig:d8}.
\end{center}
\vspace*{-0.75cm}
\end{figure} 

We obtain:
\beq
d_8= (4.3\pm 3.0)\times 10^{-2}\,{\rm GeV^8}.
\label{eq:d8-res}
\eeq
which we consider as a final result for $d_8$.  However, this low value of $d_8$ compared to the one obtained at order $\alpha_s^2$ is a good news for QCD sum rules users. It (a posteriori) justifies the neglect of high-dimension condensates in the analysis which gives a good description of different hadron parameters when only the dimension $D\leq 6$ condensates are retained in the OPE. 

From the value of $d_6$ in Eq.\,\ref{eq:res-d68}, one can deduce the value of  the four-quark condensate to order $\alpha_s^4$:
\beq
 \rho\la\bar\psi\psi\ra^2=(5.98\pm 0.64)\times 10^{-4}\,{\rm GeV^6},
 \label{eq:res-4q}
 \eeq
 which (agreably) confirms the value $(5.9\pm 0.8)\times 10^{-4}\,{\rm GeV^6}$  obtained from different determinations from light quark systems\,\cite{SNB1,SNB2,SNREV1}.
 \vspace*{-0.25cm}
\subsection*{ \b  A second attempt to extract $\la\alpha_s G^2\ra$ to order $\alpha_s^4$ }
\subsection*{\hspace*{0.25cm} \d $\la\alpha_s G^2\ra$  and $d_6$ from a two-parameter fit}
\begin{figure}[hbt]
\begin{center}
\includegraphics[width=7.5cm]{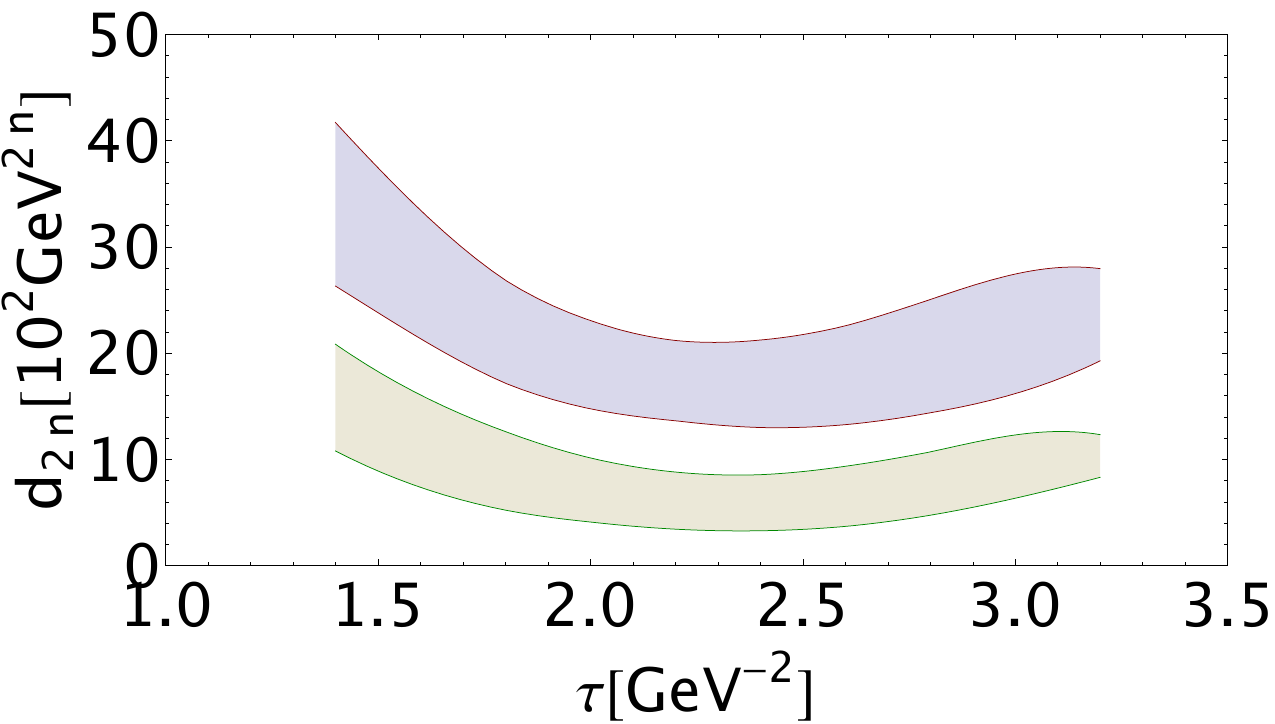}
\caption{\footnotesize  Two-parameter fit of  $\la\alpha_s G^2\ra$ (lower curves) and $d_6$ (upper curves) to order $\alpha_s^4$.} 
\label{fig:d64-as4}.
\end{center}
\vspace*{-0.75cm}
\end{figure}

First, we repeat the analysis done previously  and neglect the $d=8$ condensates. One can notice (see Fig.\,\ref{fig:d64-as4}) that the presence of the $\alpha_s^4$ term leads in this case to $\tau$-stability (minimum) at $\tau\simeq 2.4$ GeV$^{-2}$  from which we extract the optimal values:
\bea
\la\alpha_s G^2\ra &=&  (5.9\pm 2.6)\times 10^{-2}\,{\rm GeV^4},\nnb\\
 d_6&=& -(17.1\pm 4.1)\times 10^{-2}\,{\rm GeV^6},
\label{eq:d46}
\eea
where the main error is due to the data fitting procedure.
These values agree with the determination of  $\la\alpha_s G^2\ra$ from heavy quarkonia in Eq.\,\ref{eq:asg2} and with the one of $d_6$ in Eq.\,\ref{eq:res-d68} but inaccurate. 
\vspace*{-0.25cm}
\subsection*{\hspace*{0.25cm} \d $\la\alpha_s G^2\ra$  from a one-parameter fit}
\begin{figure}[hbt]
\begin{center}
\includegraphics[width=7.5cm]{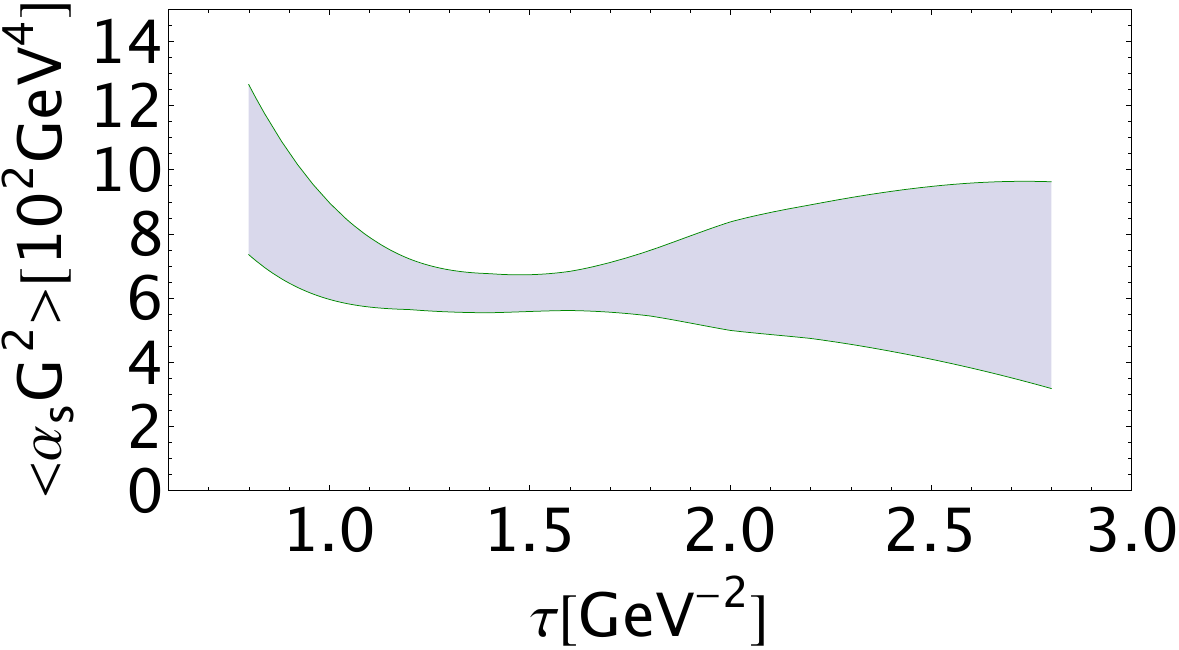}
\caption{\footnotesize  One-parameter fit of  $\la\alpha_s G^2\ra$ to order $\alpha_s^4$ using $d_6$ in Eq.\,\ref{eq:res-d68} and $d_8$ from Eq.\,\ref{eq:d8-res}.} 
\label{fig:d4}.
\end{center}
\vspace*{-0.6cm}
\end{figure}

To improve the determination of $\la\alpha_s G^2\ra$, we use as input the value of $d_6$ in Eq.\,\ref{eq:res-d68} and include $d_8$ from Eq.\,\ref{eq:d8-res}. The analysis is shown in Fig.\,\ref{fig:d4}. We deduce at the stability point $\tau\simeq 1.5$ GeV$^{-2}$:
\beq
\la\alpha_s G^2\ra =  (6.12\pm 0.61)\times 10^{-2}\,{\rm GeV^4},
\label{eq:ag2}
\eeq
which improves the result in Eq.\,\ref{eq:d46}.  Combining this value with the one from heavy quarkonia in Eq.\,\ref{eq:asg2}, we deduce the QCD spectral sum rules (QSSR) average:
\beq
\la\alpha_s G^2\ra =  (6.40\pm 0.30)\times 10^{-2}\,{\rm GeV^4}.
\eeq
\subsection*{\hspace*{0.25cm} \d Comparison with some other approaches}

\hspace*{0.25cm} -- {\it LNT\,\cite{LNT}}\,: we consider the previous  results as an improvement of  the ones of LNT (notice a slightly different normalization):
\bea
d_6\vert_{lnt} &=&  -(12.7\pm 4.2)\times 10^{-2}\,{\rm GeV^6}\, ,\nnb\\
 d_8\vert_{lnt}&=& (27\pm 24)\times 10^{-2}\,{\rm GeV^8},
\eea
 where LNT have used the same ratio of moments ${\cal R}_{10}$  with a different strategy and older data. The lower value of $d_6$ is correlated to the low value of $\la\alpha_s G^2\ra$ 
 obtained by LNT (see Fig.\,\ref{fig:d46}).
 
 \hspace*{0.25cm} -- {\it FESR\,\cite{FESR}}\,: these results are consistent in sign and in magnitude with the FESR ones:
\bea
\la\alpha_s G^2\ra\vert_{fesr}&\simeq& (7\sim 18)\times 10^{-2}~{\rm GeV}^4,\nnb\\
d_6\vert_{fesr} &=& -(0.33\sim 0.55)~{\rm GeV^6}.
\eea
The inaccuracy of the FESR results is related to their strong $t_c$-dependence 
and to the inaccuracy of the data at higher energy. 

\hspace*{0.25cm} -- {\it $\tau$-decay moments}\,:  LSR and FESR results do not favour some results with negative signs obtained from different moments of $\tau$-decay (see Table\,\ref{tab:g2}).   
\vspace*{-1cm}
   {\footnotesize
\begin{table}[hbt]
\setlength{\tabcolsep}{0.2pc}
  \begin{center}
    {
  \begin{tabular}{lllll}

&\\
\hline
\hline
$\la\alpha_s G^2\ra$&$-d_6$&$d_8$&PT & Refs.\\
 \hline 
$0.67\pm 0.89$&$15.2\pm 2.2$&$22.3\pm 2.5$&    {\rm FO}&{\rm ALEPH}\,\cite{ALEPH}\\
$5.34\pm 3.64$&$14.2\pm 3.5$&$21.3\pm 2.5$&            {\rm FO}&{\rm OPAL}\,\cite{OPAL}\\
 $0.8^{+0.7}_{-1.4}$&$32^{+8}_{-5}$&$50^{+4}_{-7}$&{\rm FO}&\,\cite{PICH1}\\
$0.31\pm 2.45$&$13.5\pm 1.8$&$20.0\pm 1.6$&       {\rm CI}&{\rm OPAL}\,\cite{OPAL}\\
$-1.57\pm0.94$&$14.7\pm 1.1$&$20.4\pm 1.3$&{\rm CI}&\,{\rm ALEPH}\,\cite{DAVIER}\\
$-0.8^{+0.7}_{-0.7}$&$35\pm 3$&$49 ^{+4}_{-5}$&  {\rm CI}&\,\cite{PICH1}\\
   \hline\hline
\end{tabular}}
 \caption{ Values of the QCD condensates of dimension $D=2n$ in units of $10^{-2}$ GeV$^{2n}$ from $\tau$-like decays \,\cite{PICH1}.  }\label{tab:g2} 
 \end{center}
\end{table}
\vspace*{-0.5cm}
} 

  \subsection*{\hspace*{0.25cm}  \d Correlated values of $d_6$ and $d_8$ versus $\la\alpha_s G^2\ra$}
\begin{figure}[hbt]
\begin{center}
\includegraphics[width=7.5cm]{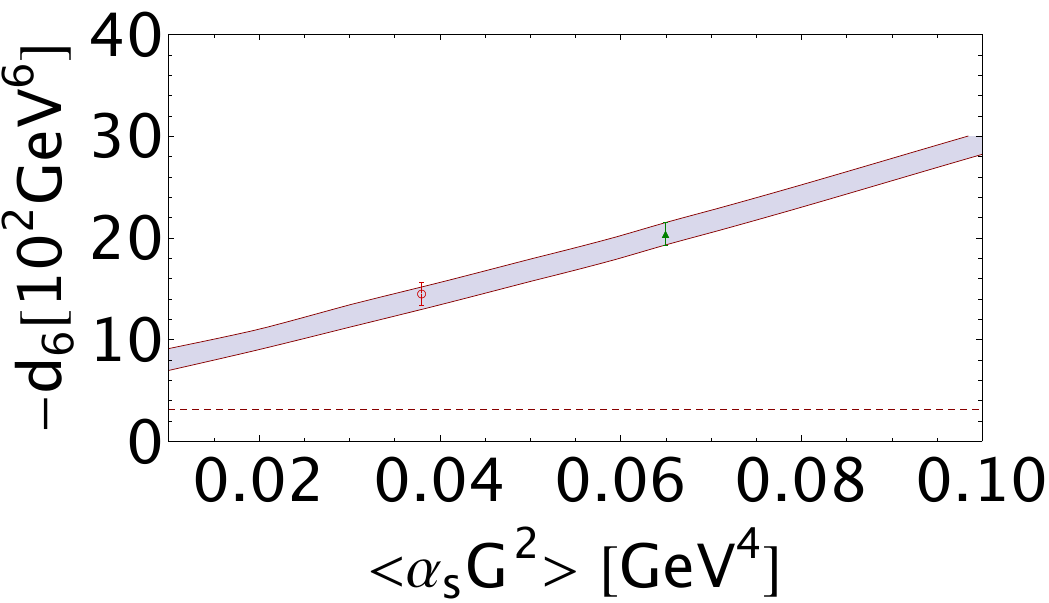}
\caption{\footnotesize  Correlated values of $d_6$ and $d_8$ versus $\la\alpha_s G^2\ra.$ The dashed horizontal line is the value of $d_6$ estimated from factorization of the four-quark condensate. The red (resp. oliva) point corresponds to the value of $\la\alpha_s G^2\ra$ given by SVZ and by Eq.\,\ref{eq:asg2}. } \label{fig:d68-g2}.
\end{center}
\vspace*{-1cm}
\end{figure}

\vspace*{-0.1cm}
We show in Fig.\,\ref{fig:d68-g2} the behaviour of $d_6$ and $d_8$ versus $\la\alpha_s G^2\ra$. One can notice that:

\hspace*{0.25cm} -- The value of the four-quark condensate estimated from factorization is inconsistent with the SVZ value of the gluon condensate.

\hspace*{0.25cm} --  The ratio:
\vspace*{-0.2cm}
\beq
r_{46} \equiv \frac{\rho\la\bar\psi\psi\ra^2}{\la\alpha_s G^2\ra}\approx 0.9\times 10^{-2}~{\rm GeV^2},
\eeq
is almost constant. This result  has also been obtained from different approaches independently on the absolute size of the extracted condensates and on the kind of methods as discussed in Ref.\,\cite{SN95}.  Moreover, we also notice that  some $\tau$-like moments results given in Table\,\ref{tab:g2} do not fulfill a such relation. 

\begin{center}
\begin{figure}[hbt]
\begin{center}
\includegraphics[width=7.5cm]{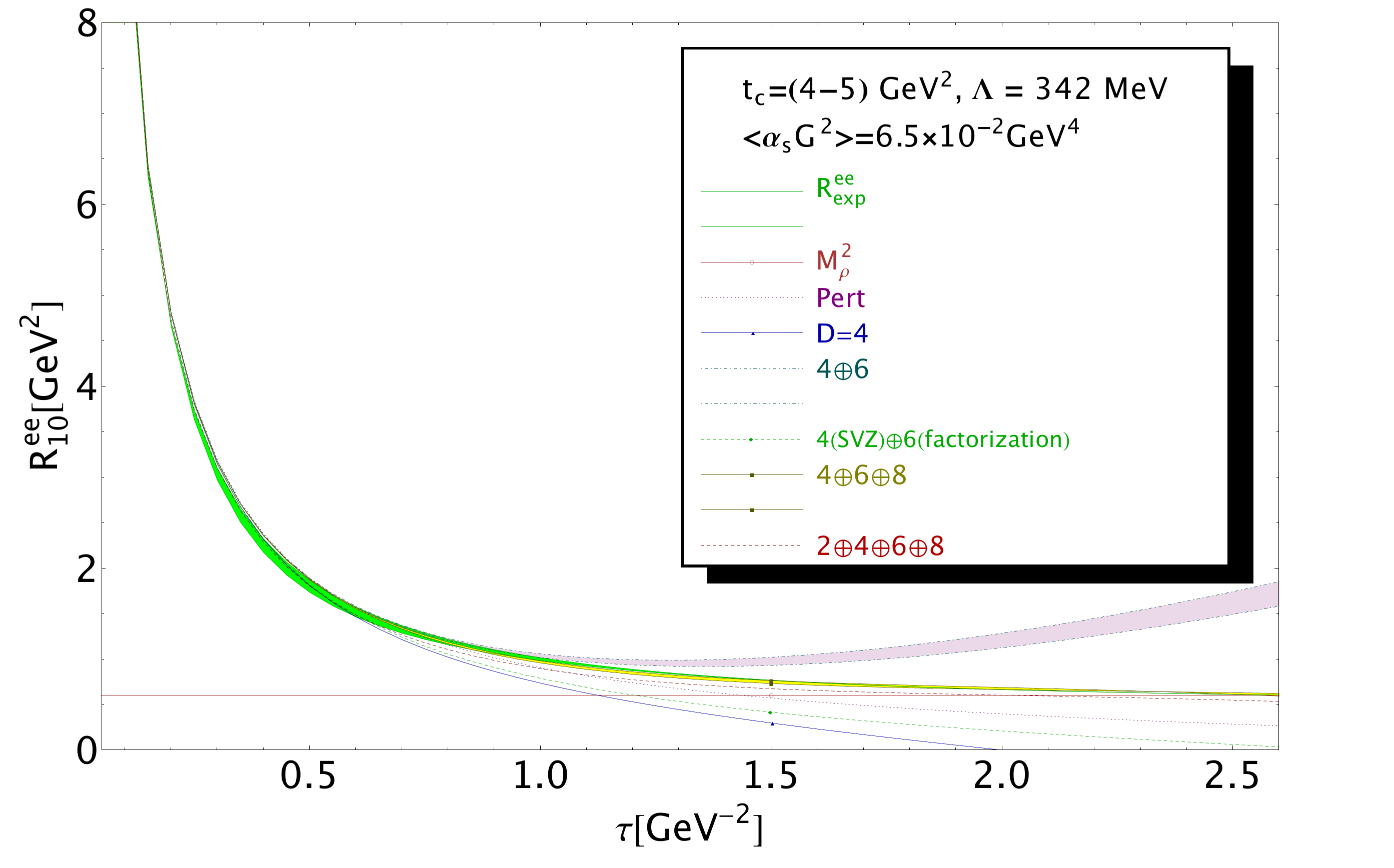}
\caption{\footnotesize  $\tau$-behaviour of the ratio of moments ${\cal R}^{ee}_{10}$ for different truncation of the OPE.}\label{fig:Ree}.
\end{center}
\vspace*{-0.75cm}
\end{figure} 
\end{center}
\subsection*{\b Confronting theory and experiment for ${\cal R}^{ee}_{10}$ and ${\cal L}^{ee}_0$}
Using as input the previous values of the condensates and the value of $\alpha_s$ from PDG\,\cite{PDG}, we compare the QCD and experimental sides of  ${\cal R}_{10}$ and ${\cal L}_0$ which we show in Figs.\,\ref{fig:Ree} and \ref{fig:Lee}.
  \subsection*{\hspace*{0.25cm}  \d Ratio of moments ${\cal R}_{10}$}
\hspace*{0.25cm}-- The use of the standard SVZ value of the gluon condensate $\oplus$ the value of the four-quark condensate estimated using the factorization assumption underestimates the data. However, the simultaneous use of the SVZ standard value $\oplus$ the factorization (combination often used in the QCD sum rules literature) is inconsistent from the result of simultaneous two-parameter fit   shown in the previous section and in Fig.\,\ref{fig:d68-g2}.

\hspace*{0.25cm} -- The inclusion of the $D=8$ condensate enlarges the region of agreement between the QCD prediction and the experiment until $\tau=2.5$ GeV$^{-2}$ where the $\rho$-meson mass is reached and where, exceptionally, the PT series for the spectral function still make sense. 

\hspace*{0.25cm} -- We complete the analysis by adding the contribution of the dimension 2 tachyonic gluon mass beyond the SVZ-expansion\,\cite{CNZa}. We see that it tends to decrease the agreement with the data but the effect is (almost) negligible within our precision in the high-energy region. 
\begin{center}
\vspace*{-0.4cm}
\begin{figure}[hbt]
\begin{center}
\includegraphics[width=7.5cm]{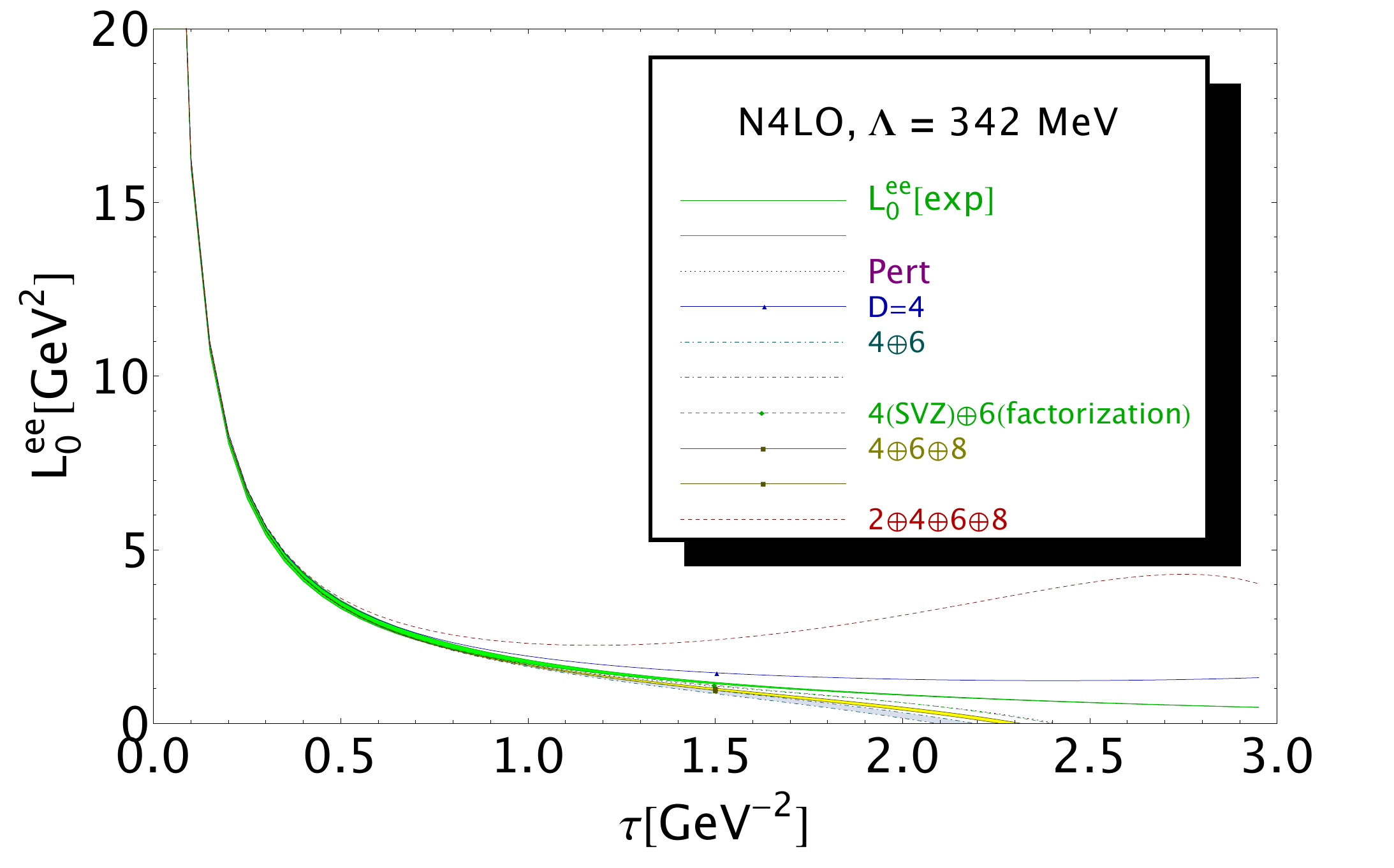}
\caption{\footnotesize  $\tau$-behaviour of the lowest moment ${\cal L}_0$ for different truncation of the OPE.}\label{fig:Lee}
\end{center}
\vspace*{-0.75cm}
\end{figure} 
\end{center}
\subsection*{\hspace*{0.25cm}  \d Lowest moment ${\cal L}^{ee}_{0}$}
 We have tried to extract these parameters using ${\cal L}^{ee}_{0}$ but the results are unconclusive. 
 
\hspace*{0.25cm} --  Contrary to the case of 
${\cal R}^{ee}_{10}$, we notice an important sensitivity of the results on the truncation of the PT series. This problem might be evaded if one uses the $\pi^2$ resummation of the higher order terms where the PT series converges faster\,\cite{KAHN}. 

\hspace*{0.25cm} -- However, we also remark that the determination from the low and high sets of data points varies in a large range leading to a very inaccurate result. 
  \section{The lowest $\tau$-like decay moment}
 The lowset $\tau$-like decay moment applied to $e^+e^-\to $ Hadrons data have been initially used in Refs.\,\cite{SNPICH,SN95} to extract $\alpha_s$ and the QCD condensates. It reads\,\cite{BNP}:
  \beq
  R^{ee}_{\tau}=\int_0^{1} dx_0\, (1-3x_0^2+2x_0^3)2R^{I=1}_{ee}(x_0),
  \eeq
  with $x_0\equiv (t/M^2_0)$. We attempt to extract the value of $\alpha_s$ for different values of $M_0$. We use the PT  expression up to the calculated order $\alpha_s^4$ and estimate the $\alpha_s^5$ term assuming a geometric growth of the PT coefficients:
   \beq
 p_5^ {FO}\approx +597, ~~~~~~~ p_5^{CI}\approx +426~.
  \label{eq:as5}
  \eeq
   We include into the OPE the contributions of the condensates up to dimension 8 and do not consider non-standard ones (tachyonic gluon, instantons, duality violation) which have been estimated to be small\,\cite{SN95,SNTAU}.  
  \subsection*{\b Extraction of $\alpha_s(M_\tau)$}  
 Using the previous values of the QCD non-perturbative (NP) parameters, the values of $\alpha_s(M_\tau)$ for different  $M_0$ are shown in Fig.\,\ref{fig:as} for Fixed Order (FO) and Contour Improved (CI) PT series. We obtain to order $\alpha_s^4$\,:
   \bea
\hspace*{-0.5cm}  \alpha_s^{FO}(M_\tau)&=&0.3247 (46)_{fit}(1)_{np} (62)_{h.o}\nnb\\&=&0.3247 (77)~~~\lrar2 \hspace*{-0.45cm\circ}\nnb\\ \alpha_s^{FO}(M_Z)&=&0.1191(10)(3)\, \nnb\\
   \alpha_s^{CI}(M_\tau)&=&0.3483 (63)_{fit}(1)_{np} (15)_{h.o}\nnb\\
   &=&0.3483 (65)~~~\lrar2 \hspace*{-0.45cm\circ}\nnb\\ \alpha_s^{CI}(M_Z)&=&0.1218(8)(3)\, ,
  \label{eq:alphas}
  \eea
 where the dominant error comes from the fitting procedure and the estimate of the higher order corrections.  The last error in $\alpha_s(M_Z)$ comes from the running procedure. 
  We deduce the (conservative) average to order $\alpha_s^4$:
  \bea
 \la \alpha_s(M_\tau)\ra &=&0.3336 (31)(116)_{syst}~~~\lrar2 \hspace*{-0.45cm\circ}\nnb\\
 \la  \alpha_s(M_Z)\ra&=&0.1201(14)(3)\, .
  \eea
  where the first error is due to the fitting procedure, while the systematic error is the distance between the mean and the central FO/CI values. $\pm 3 $ is
  an error induced by the running procedure.  If one takes literally the estimate of the $\alpha_s^5$ coefficients in Eq.\,\ref{eq:as5}, the result becomes\,:
  \bea
 \la \alpha_s(M_\tau)\ra &=&0.3263(37)(78)_{syst}~~~\lrar2 \hspace*{-0.45cm\circ}\nnb\\
 \la  \alpha_s(M_Z)\ra&=&0.1193(11)(3)\,. 
  \eea
\begin{center}
\vspace*{-0.4cm}
\begin{figure}[hbt]
\begin{center}
\includegraphics[width=7.5cm]{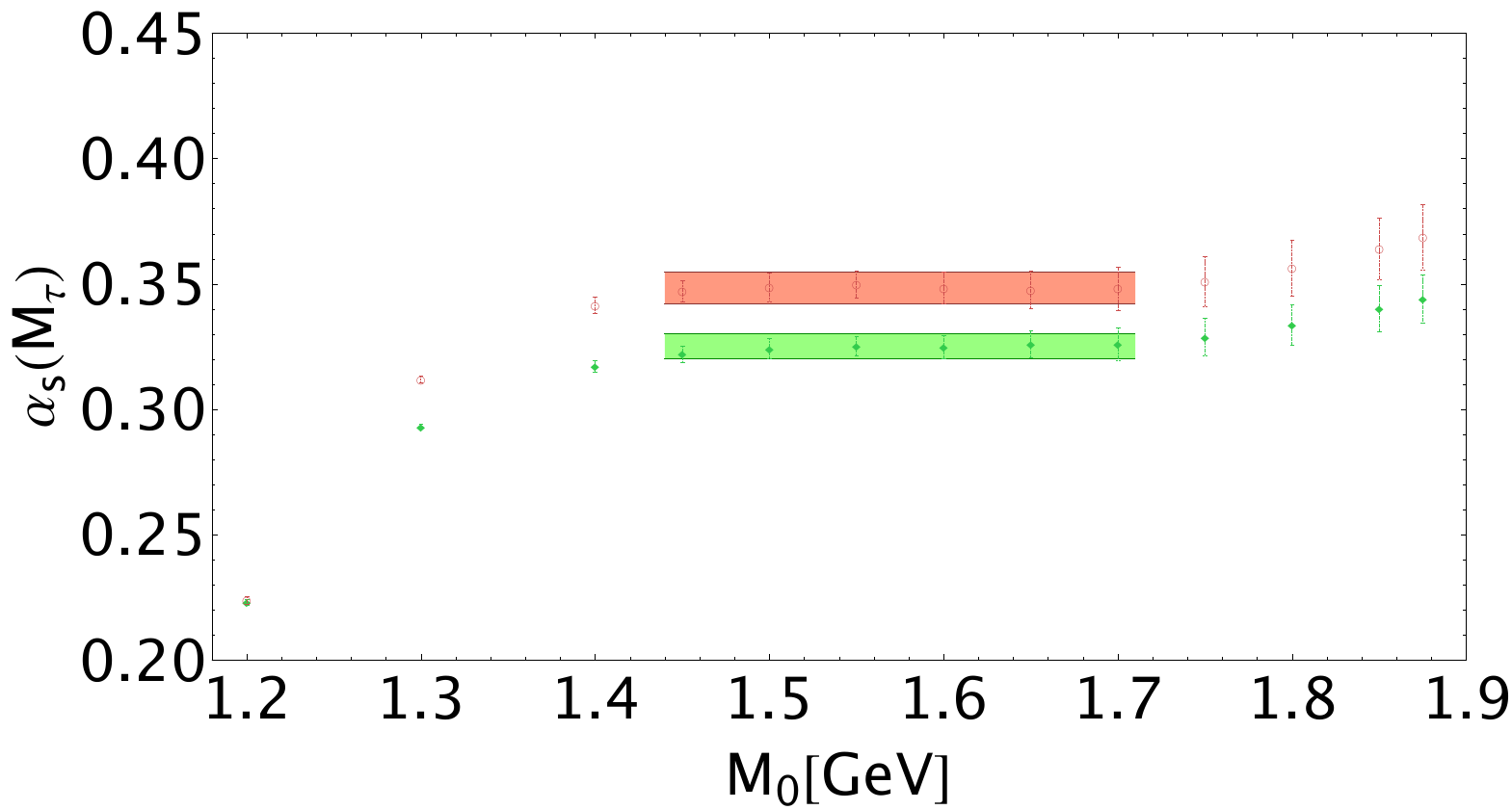}
\caption{\footnotesize Value of $\alpha_s(M_\tau)$ to ${\cal O}(\alpha_s^4)$ as a function of the hypothetical $\tau$-mass $M_0$ including $D\leq 8$ condensates. Green : FO. Red : CI. The errors due to the non-perturbative condensates and to the fitting procedure are shown.}\label{fig:as}.
\end{center}
\vspace*{-0.75cm}
\end{figure} 
\end{center}
   {
   \scriptsize
\begin{table}[hbt]
\setlength{\tabcolsep}{0.1pc}
  \begin{center}
    {
  \begin{tabular}{lllll}

&\\
\hline
\hline
$\alpha_s ^{FO}(M_\tau)$&$\alpha_s ^{CI}(M_\tau)$&$\delta^V_{NP}(M_\tau)\times 10^{-2}$  &Data& Refs.\\
 \hline 
\boldmath$0.3247(77)$&\boldmath$0.3483(65)$&\boldmath$2.30\pm 0.20$&\boldmath$e^+e^-$&{\bf This}\\
0.320(30)&--& $3.60\pm 1.64$&  $e^+e^-\oplus$ $\tau$ & \cite{SN95}\\
0.350(50)&--&$1.45\pm 1.3$ &$e^+e^-$& \cite{SNPICH} \\
0.320(22)&0.340(23)&$2.0\pm1.7$&V\,: $\tau$&\cite{ALEPH}\\
0.323(16)&0.347(23)&$1.87\pm 0.54$&V\,: $\tau$&\cite{OPAL}\\
0.328(9)&--&--& V+A\,:  $\tau$ & \cite{SNTAU}\\
0.322(16)&0.342(16)&-- &V+A\,: $\tau$&\cite{CHET4}\\
0.324(11)$^{*)}$&0.346(11) &--&V+A\,: $\tau$&\cite{DAVIER}\\
0.320(12)&0.335(13)&--&V+A\,: $\tau$&\cite{PICH2}\\
\hline
\it  0.3243(42) &\it 0.3452(47)&$\it 2.28\pm 0.20$&& \it Mean \\
   \hline\hline
\end{tabular}}
 \caption{ ~$\alpha_s(M_\tau)$ and $\delta^V_{NP}$ from $\tau$-decay moments within fixed order (FO) and contour improved (CI) perturbative series up to ${\cal O}(\alpha_s^4)$. $^{*)}$ indicates that the quoted error is our crude estimate. }\label{tab:alpha_s} 
 \end{center}
   \vspace*{-0.75cm}
\end{table}
} 
  \subsection*{\b Power corrections}
\subsection*{\hspace*{0.25cm}\d $d_6$  from a one-parameter fit}
\begin{figure}[hbt]
\begin{center}
\includegraphics[width=7.5cm]{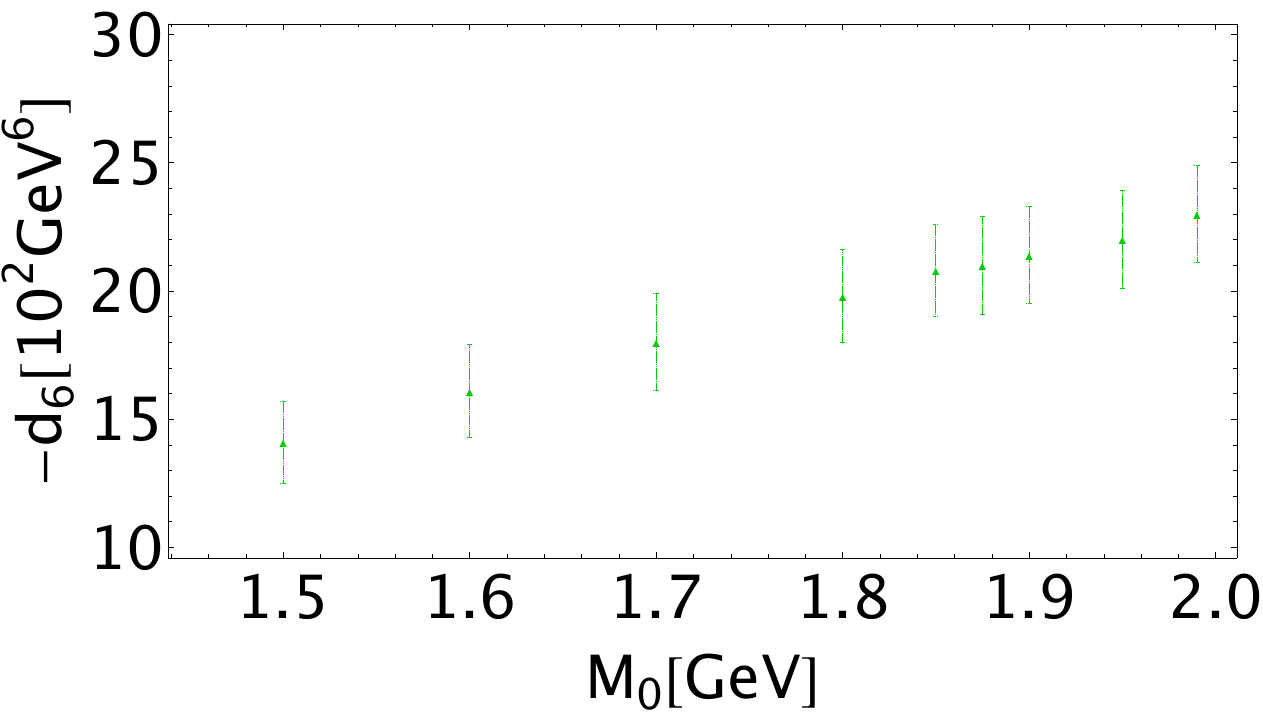}
\caption{\footnotesize $d_6$ condensate as function of the hypothetical $\tau$-mass $M_0$ using $\alpha_s(M_0), d_4$ and $d_8$ from ${\cal R}_{10}$ as inputs.}\label{fig:d6-tau}.
\end{center}
\vspace*{-0.75cm}
\end{figure}

To give a stronger constraint on $d_6$, we shall work with a one-parameter fit.   We shall use as input the value of the gluon condensate in Eq.\,\ref{eq:asg2} and the previously determined values of $\alpha_s(M_\tau)$ from $R^{ee}_{\tau}$ in Eq.\,\ref{eq:alphas}   and  of $d_8$ in Eq.\ref{eq:res-d68} from ${\cal R}_{10}$. 
The result of the analysis is shown in Fig.\,\ref{fig:d6-tau} where one remarks that the result increases softly with the value of $M_0$ and shows  an almost stability (inflection point) around (1.8-1.9) GeV which is: 
\beq
d_6\simeq -(20\sim 21)\times 10^{-2}~{\rm GeV}^6
\eeq
with a 10\% error. This result is identical to the one in Eq.\,\ref{eq:res-d68}. However, the results obtained from some other $\tau$-decay moments are also recovered in the $M_0$-instability region.  Then, we conclude that the extraction of $d_6$ from the low $\tau$-decay moment is less accurate than the one from $R_{10}^{ee}$. It only gives an approximate range of values. 


  \subsection*{\hspace*{0.25cm}\d  Sum of power corrections to the lowest moment $R^{ee}_{\tau}$}
  Using our previous determinations of the condensates, we can estimate the sum of the power corrections to $R^{ee}_{\tau}$. In so doing, we introduce as input the value of $\alpha_s(M_\tau)$ determined previously and the value of the gluon condensate in Eq.\,\ref{eq:asg2}. 
  
 \hspace*{0.5cm}-- To minimize the number of free parameters, we first neglect the contribution of high-dimension condensates $D\geq 10$ and assume that the retained condensate $D=8$ is an effective condensate which absorbs into it all unknown high-dimension condensate effects. Then, using the values of $d_4$ in Eq.\ref{eq:asg2}, $d_6$ and $d_8$ in Eq.\,\ref{eq:res-d68}, we deduce the sum of the NP contributions\,:
\beq
\delta^V_{NP}(M_\tau)= (2.3\pm 0.2)\times 10^{-2},
\eeq
which improves our previous findings from $e^+e^-\to I=1$ hadrons data\,: $\delta^V_{NP}(M_\tau)= (2.38\pm 0.89)\times 10^{-2}$ in Ref.\,\cite{SN95} and the analysis in\,\cite{SNTAU}.  

\hspace*{0.5cm}-- It is remarkable to notice that this value agrees with the one from $\tau$-decay analysis (see Table\,\ref{tab:alpha_s}) despite the large discrepancies with the individual values of each condensate.  This is due to the alternate signs of the condensate contributions in the $\tau$-moments.

\hspace*{0.5cm}-- One can also notice that the value of $\alpha_s$ decreases with $\delta_{NP}$.  From $\delta^V_{NP}=3.7\times 10^{-2}$  (our first iteration) to the final value: $2.3 \times 10^{-2}$ (Eqs.\,\ref{eq:res-d68} and \ref{eq:d8-res}), the value of $\alpha_s(M_\tau)$ moves slightly  from 0.329 to 0.325. 
\vspace*{-0.25cm}
\section{$a_{\mu,\tau}\vert^{hvp}_{l.o}$ from   $e^+e^-\to$ hadrons data}
We complete the analysis by updating our previous determination of the  lowest order hadronic contributions  to the vacuum polarization of $a_{\mu,\tau}$\,\cite{CALMET,SN76,SNamu,SN78}.  This analysis will also serve as a test of our parametrization of the $I=1$ part of the spectral function used in previous sections especially in the $\rho$ meson region. 

The lowest order hadronic contributions to the vacuum polarization of a lepton $l\equiv \mu,\tau$ can be obtained from the
well-known dispersion relation\,\cite{BOUCHIAT,DURAND,KINOSHITA,BOWCOCK,GOURDIN}\,:
\beq
a_l\equiv \frac{1}{2}(g-2)_l= \frac{1}{4\pi^3}\int_{4m_\pi^2}^\infty dt\, K_l(t)\sigma (e^+e^-\to {\rm hadrons})
\eeq
where $K_l(t)$ is the QED kernel function\,\cite{LAUTRUP}:
\beq
K_l(t)=\int_0^1dx\frac{x^2(1-x)}{x^2+(t/m_l^2)(1-x)}.
\label{eq:kmu}
\eeq
This analysis is an update of the one in Refs.\,\cite{CALMET,SN76,SNamu}.
\subsection*{\b Light $I=1$ mesons below 1.875 GeV}
 For the pion form factor below 0.993 GeV,  instead of the usual Breit-Wigner parametrization, we subdivise this region  into six subregions: [$2m_\pi$, 0.5] tail of the Breit-Wigner and polynomial fits for [0.50,0.60], [0.60, 0.778], [0.778, 0.786], [0.786, 0.810] and [0.810, 0.993] GeV.  Above 0.993 GeV, we use the same parametrization as in the previous section. 
 In the region between 0.6 to 0.88 GeV, we obtain using the CMD3 data\,\cite{CMD3}\,:  
$
a_\mu^{\rho}\vert^{hvp}_{l.o}[0.6\to 0.88]=(377.4\pm 3.1)\,\times 10^{-10},
$
which agrees with the CMD3  estimate but higher than the KLOE one by   $16.8\,\times 10^{-10}$ (see Fig. 17 of  Ref.1 for  estimates using some other data). 

\subsection*{\b Light $I=0$ mesons  below 1.875 GeV}
We use NWA for the $\omega$ and $\phi$ mesons. We use a Breit-Wigner with the mass and widths given by PDG  for the $\omega(1650)$ and $\phi(1680)$. mesons 
\subsection*{\b Light $I=0\oplus 1$ mesons from 1.875 t0 3.68 GeV}
\subsection*{\hspace*{0.25cm} \d From 1.875 to 2 GeV}
We subdivide this region into three ones [1.875,1.91], [1.91,1.96], [1.96,2] GeV.  where we use polynomial fits 
of the data.
\subsection*{\hspace*{0.25cm} \d From 2 to 3.68 GeV}
The data in this region are well fitted by the QCD expression of the spectral function for 3 flavours as one can see from Fig.\,\ref{fig:charm} given by PDG. 
To the massless PT expression known to order $\alpha_s^4$, we  include the quark and gluon condensates of dimensions $D=4,6$.  We add the quadratic  $m^2_s$-corrections to order $\alpha_s^3$ and the quartic mass $m_s^4$ corrections to $\alpha_s^2$.  We use the Renormalization Group Invariant (RGI) mass : $\hat m_s$=114(6) MeV\,\cite{SNREV1}. 
\subsection*{\b  Charmonium}
\subsection*{\hspace*{0.25cm} \d The $J/\psi(1S),\psi(2S)$ and $\psi(3773)$}
We estimate their contributions using a NWA and the values of the masses and widths from PDG\,\cite{PDG}:
\subsection*{\hspace*{0.25cm} \d From 3.68 to 4.55 GeV  }
We divide this region into 5 subregions\,: [3.68, 3.86], [3.86,4.10], [4.10,4.18], [4.18,4.30] and [4.30,5.55] GeV. The region between 3.68 to 3.86 is better fitted using a Breit-Wigner while in the others,  we use polynomials.
\subsection*{\hspace*{0.25cm}\d From 4.55 to 10.50 GeV }
\begin{figure}[hbt]
\begin{center}
\includegraphics[width=7.5cm]{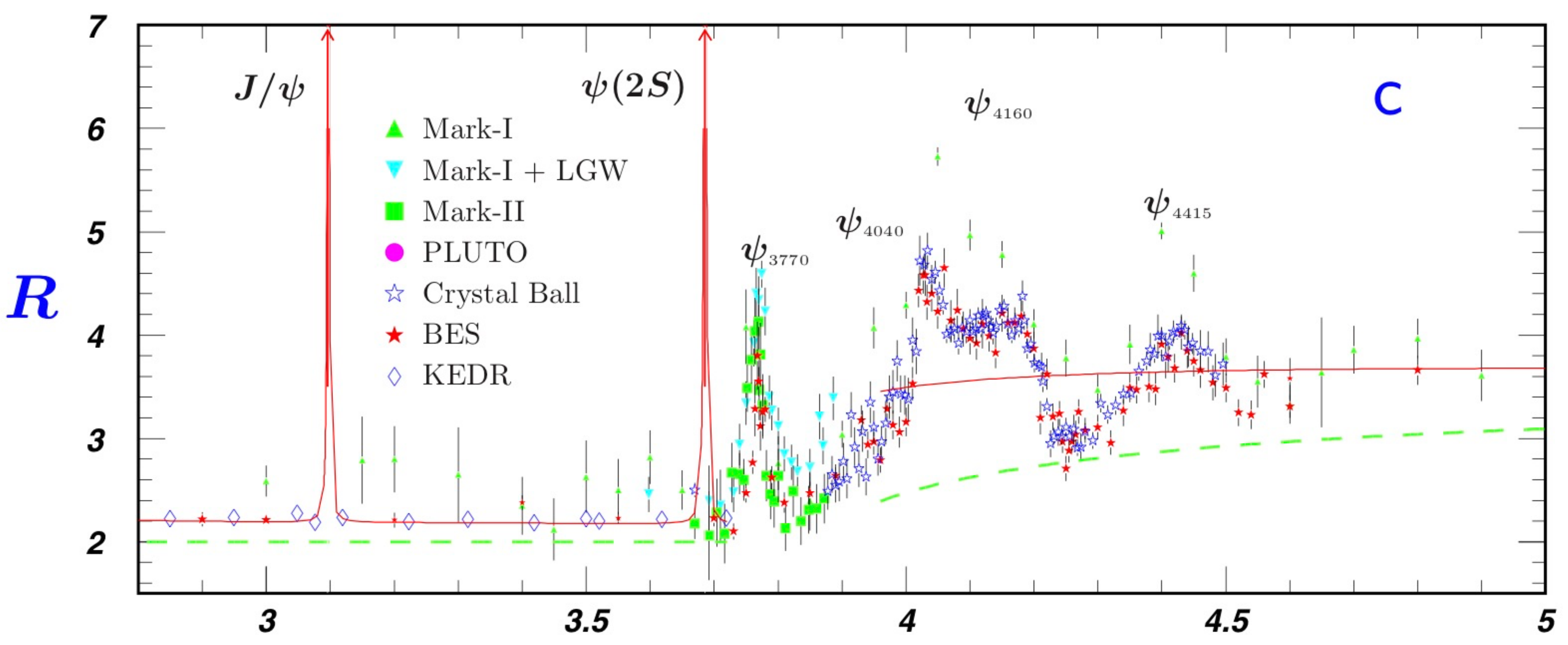}
\caption{\footnotesize  $e^+e^-\to$ Hadrons data in the charmonium region from 2 to 5 GeV from PDG\,\cite{PDG}. } \label{fig:charm}
\end{center}
\vspace*{-0.75cm}
\end{figure}

\begin{figure}[hbt]
\begin{center}
\includegraphics[width=7.5cm]{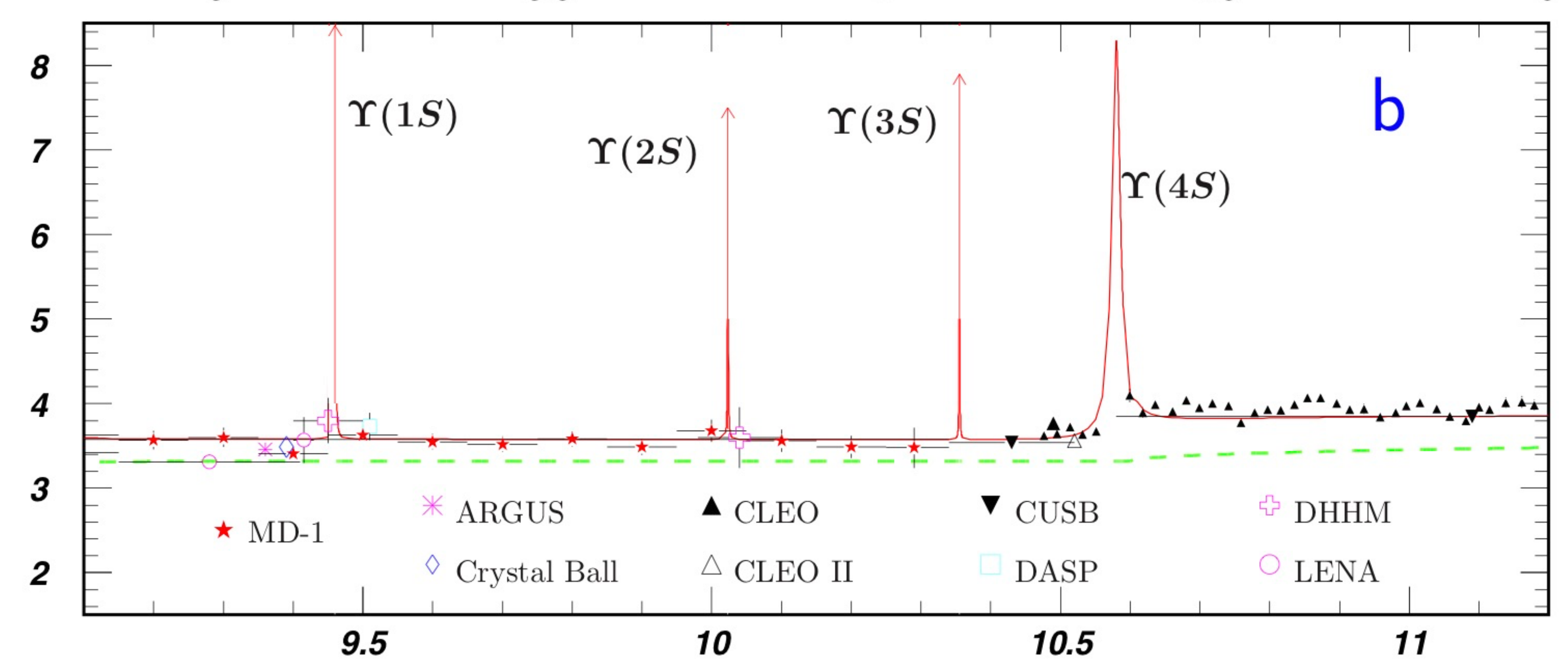}
\caption{\footnotesize  $e^+e^-\to$ Hadrons data in the bottomium region from 9 to 10.5 GeV from PDG\,\cite{PDG}. } \label{fig:bottom}
\end{center}
\vspace*{-0.75cm}
\end{figure}

As shown in Figs.\,\ref{fig:charm} and \ref{fig:bottom}, the data without resonance peaks are well fitted by QCD for 4 flavours. We add to the previous QCD expressions the charm contributions with $m_c^2$ and $m_c^4$ mass corrections where $m_c(m_c)=1266(6)$ MeV\,\cite{SNREV1}.

\subsection*{\b   Bottomium}
\subsection*{ \hspace*{0.25cm} \d $\Upsilon(1S\to 11.02)$}
We use a NWA to estimate their contributions and use the masses and leptonic widths given by PDG.

\subsection*{ \hspace*{0.25cm} \d From  10.59 GeV to $2m_t$}
We parametrize the spectral function using the QCD continuum. 
We add the $b$-quark contribution to the previous QCD expression where $b$-quark mass corrections to order $a_s^3\overline{m}_b^2/t$ and $a_s^2\overline{m}_b^4/t^2$ are included. We use $m_b(m_b)=4202(8)$ MeV\,\cite{SNREV1}.
For the analysis, we consider the region from 10.59 MeV to $2m_t$ just after the $\Upsilon(4S)$ where the QCD continuum is expected to smear the  $\Upsilon(10860, 11020)$  and some eventual  higher resonances. 
\subsection*{ \vspace*{0.5cm} \d QCD continuum contribution from  $2m_t\to\infty$}
\vspace*{-0.25cm}
Due to the heaviness of the top quark mass, we shall use the approximate Schwinger formula near the $\bar tt$ threshold
for a much better description of the spectral function up to order $\alpha_s$ due to the top quark:
\vspace*{-0.25cm}
\beq
R_t^{ee}=\frac{4}{3} v\frac{(3-v^2)}{2}\Big{[} 1+\frac{4}{3}\alpha_s\,f(v)\Big{]},
\eeq
\vspace*{-0.25cm}
with:
\vspace*{-0.15cm}
\beq
f(v)=\frac{\pi}{2\, v}-\frac{(3+v)}{4}\ga \frac{\pi}{2}-\frac{3}{4\pi}\dr\,:\,\,\,\,\,\, v=\ga 1-\frac{m_t^2}{t}\dr^{1/2}.
\eeq
Here $m_t$ is the on-shell top quark mass which we fix to be\,\cite{PDG}\,\footnote{We should note that the definition of the top quark mass from different experiments is still ambiguous.}: 
\beq
\overline{m}_t(\overline{m}_t^2)=(172.7\pm 0.3)\lrar2 \hspace*{-0.45cm\circ}~~~~
\hat m_t=(254\pm 0.4)\,{\rm GeV},
\label{eq:mt}
\eeq
 from some  direct measurements while $\hat m_t$ is the RGI top mass defined in  Eq.\,\ref{eq:mt}.
We add to this expression the  one due to $\alpha_s^2$ and $\alpha_s^3$ within the $\overline {MS}$-scheme. Adding the above expressions to the ones in the previous sections, we obtain the result given in Table\,\ref{tab:amu1}. 
\subsection*{\b Conclusion for $a_\mu\vert^{hvp}_{l.o}$} 
We have used  the sum  exclusive of $e^+e^-\to$ hadrons data compiled by PDG $\oplus$ some resonances to extract the 
lowest order vacuum polarization contribution to the muon anomaly. We find from Table\,\ref{tab:amu1}\,:
\beq
a_\mu\vert^{hvp}_{l.o}=(7036.5\pm 38.9)\times 10^{-11},
\eeq
where the largest contribution and error come (as expected) from the $\rho$-meson low-energy one. It is amazing that the central value remains stable when comparing it with our old results in Refs.\,\cite{CALMET,SN76,SNamu} though the accuracy has increased by a huge factor of about 24\,(!) thanks to the experimental efforts for improving the data during about half century !  The total sum is slightly larger  than the recent data based determinations in Refs.\,\cite{NOMURA,DAVIER2}\,\footnote{Some earlier references are quoted in these papers and in\,\cite{SNamu}.} but in better agreement with a recent analysis of $\tau$-decay data\,\cite{MIRANDA}  and with some recent lattice results\,\cite{RAF,GM2,LATTICE}.  Using this value into Table 8 of Ref.\,\cite{KNECHT} and in Table 1 of Ref.\,\cite{GM2} where some other sources of contributions are reviewed, we deduce:
\beq
a_\mu^{th}= 116591916(42)\times 10^{-11}.
\eeq
This leads to:
\beq
\Delta a_\mu\equiv a_\mu^{exp}-a_\mu^{th} = (143\pm 42_{th}\pm 22_{exp})\times 10^{-11},
\label{eq:amu-sm}
\eeq
which indicates about $3\,\sigma$ discrepancy between experiment and the SM predictions. We have used the new experimental data\,\cite{MG2}\,:
\beq
a_\mu^{exp}= 116592059(22)\times 10^{-11},
\eeq
which improves the accuracy of previous  results in Refs.\,\cite{BNL} and FNAL\,\cite{FNAL} \, by a factor 1.86. 

\subsection*{\b Some comments on the determination of  $a_\mu\vert^{hvp}_{l.o}$} 

\d Our analysis differs from the most recent ones in Refs.\,\cite{NOMURA,DAVIER2} as we fit the sum of exclusive modes except the $\rho$-meson which we have subdivided into subregions. For the narrow resonances we use the NWA.

\d Compared to the existing analysis, we have also studied in details the contributions from the heavy quark sectors taking into account all possible resonances and, in particular, analyzed carefully the charmonium region. 

\d We have also carefully parametrized the QCD continuum contributions taking into account higher order PT quark mass corrections and the non-perturbative ones. 

\d Comparing our results with the ones in Refs.\,\cite{DAVIER2,NOMURA}, we found that in the low-energy region below 1.875 GeV, our result is higher by about $(100-136)\times 10^{-11}$ than the ones in these references. This is mainly due to the pion form factor where we use the new data of CMD3 $\oplus$ PDG which leads to higher value of $a_\mu$  than some other determinations in the region $0.60\leq \sqrt{t}\leq 0.88$ GeV (see Fig. 17 of Ref.\cite{SNe}).

\d In the high-energy region $\sqrt{t}\geq 1.875$ GeV, our result is $(545.3\pm 2.2)\times 10^{-11}$ which is  about the same as  the one $(535.5\pm 7.0)\times 10^{-11}$ of Ref.\,\cite{NOMURA}  but smaller by about 38$\times 10^{-11}$ than $(583.6\pm 3.3)\times 10^{-11}$ in Ref.\,\cite{DAVIER2}. This difference is mainly due to  the choice of the QCD continuum threshold $\sqrt{t_c}= 1.8$ GeV in Ref.\,\cite{DAVIER2}  which is lower than the one found from the asymptotic coincidence of the two sides of ${\cal L}_0$ Laplace moment  in Eq.\,\ref{eq:tc} and requested by duality from FESR. 

\subsection*{\b Extension to  $a_\tau\vert^{hvp}_{l.o}$} 
\d We extend the muon analysis to the case of the $\tau$-lepton by simply changing the lepton mass. The total sum is from Table\,\ref{tab:amu1}\,:
\beq
a_\tau\vert^{hvp}_{l.o}=(3494.8\pm 24.7)\times 10^{-9},
\eeq
which we consider as an improvement of the {\it pioneer determination} of this quantity and of $a_\tau$ in Ref.\,\cite{SN78}. 

\d One can notice that the relative weight of the light quarks over the heavy ones moves from 23 to 6 from $\mu$ to $\tau$ indicating that a measurement of the $\tau$-anomaly will probe higher energy region thanks to the behaviour of the QED kernel function  given in Eq.\,\ref{eq:kmu}. 

\d Our result is slightly lower than the one of Ref.\,\cite{NOMURA} where the origin may come from the $I=0$ light mesons region. 
\vspace*{-0.25cm}
\section{Determination of $ \Delta \alpha_{had}^{(5)} (M_Z^2)$}
We conclude the paper by updating our previous determination of $\Delta \alpha_{had}^{(5)} (M_Z^2)$ in Ref.\,\cite{SNalfa}. The hadronic contribution to this quantity can be expressed as:
\beq
\Delta \alpha_{had}^{(5)} (M_Z^2)=-\ga \frac{\alpha}{3\pi}\dr M^2_Z\int_{4m_\pi^2}^\infty \frac{R_{ee}(t)}{t(t-M_Z^2)},
\eeq
where $R_{ee}$ is the ratio of the $e^+e^-\to$ Hadrons over the  $e^+e^-\to \mu^+\mu^-$ total cross-sections. 
The results from different regions are shown in Table\,\ref{tab:amu1} where at the  $Z_0$ pole, we take the principal value of the integral which we take as:
\beq
\int_{4m_\pi^2}^{(M_Z- \Gamma_Z/2)^2}\hspace*{-1cm} dt\,f(t)+\int_{(M_Z+ \Gamma_Z/2)^2}^{\infty}\hspace*{-1cm}dt\,f(t),
\eeq
where $\Gamma =2.5$ GeV is the total hadronic $Z$-width. We add to the QCD continuum contribution the one of the $Z$-pole estimated to be\,\cite{YND17}:
$
\Delta \alpha_{had}^{(5)}\vert_{M_Z}= 29.2\times 10^{-5}.
$
Then, we obtain the total sum:
\beq
\Delta \alpha_{had}^{(5)} (M_Z^2)= (2766.3\pm 4.5)\times 10^{-5}. 
\eeq 
This value is comparable with the ones in the literature reviewed e.g. in Ref.\,\cite{JEGER}.  It improves and confirms our previous determination in Ref.\,\cite{SNalfa}. 

   {\footnotesize
\begin{table*}[hbt]
\setlength{\tabcolsep}{1.7pc}
  \begin{center}
    { \footnotesize
  \begin{tabular}{| llll |}
\hline
\hline
\rowcolor{yellow}\boldmath$\sqrt{t}$\,\bf [GeV] &\boldmath$a_\mu\vert^{hvp}_{l.o}\times 10^{11} $&\boldmath$a_\tau\vert^{hvp}_{l.o}\times 10^{9} $&$\Delta\alpha^{(5)}_{had}(M_Z^2)\times 10^5$\\ 
 \hline 

{\bf Light I=1}&&&\\
$\rho(2m_\pi\to 0.50)$&$489.4\pm 3.1$&$85.1\pm 0.6$&$12.38\pm 0.08$\\
$\rho(0.50\to 0.60)$&$524.5\pm 13.7$&$135.9\pm 3.5$&$22.84\pm 0.59$\\
$\rho(0.60\to 0.776)$&$2712.0\pm 30.2$&$943.7\pm10.3$&$182.87\pm 1.97$ \\

$\rho(0.776\to 0.993)$&$1297.3\pm9.8$&$1797.3\pm 20.3$&$117.82\pm 0.98$\\
$0.993\to 1.5$&$354.4\pm 6.7$&$228.2\pm 4.1$&$67.3\pm 1.14$\\
$1.5\to 1.875$&$237.6\pm 5.7$&$206.8\pm 4.9$&$80.11\pm 1.9$\\

{\it Total Light I=1 ($\leq 1.875$)}&$\it 5615.2\pm 36.0$&$\it 2453.3\pm 11.9$&$\it 483.3\pm 4.4$\\
{\bf Light I=0}&&&\\
$\omega$ (NWA) &$417.1\pm 13.7$&$163.3\pm 5.3$&$33.6\pm 1.1$\\
$\phi$ (NWA) &$389.6\pm 4.6$&$20.6\pm 0.2$&$51.2\pm 0.6$\\
$0.993\to 1.5$ & $44.3\pm 0.8$&$28.5\pm 0.5$&$8.4\pm 0.1$\\
$\omega(1650)$(BW)&$24.3\pm 0.1$&$16.7\pm 0.1$&$5.2\pm 0.1$ \\
$\phi(1680)$(BW)&$1.8\pm 0.9$&$1.3\pm 0.6$&$0.4\pm 0.2$ \\
{\it Total Light I=0 ($\leq 1.875$)}&$\it 877.1\pm 14.5$&$\it 230.4\pm 5.3$&$\it 99.4\pm 1.3$\\
{\bf Light I=\boldmath $0\, \oplus \,1$}&&&\\
$1.875\to 1.913$ & $14.7\pm 0.7$&$14.4\pm 0.7$&$6.3\pm 0.3$\\
$1.913\to 1.96$ & $17.6\pm 0.6$&$17.5\pm 0.6$&$7.9\pm 0.3$\\
$1.96\to 2$ & $14.3\pm 0.5$&$14.5\pm 0.5$&$6.7\pm 0.2$\\
$2\to 3.68$:  QCD $(u,d,s)$ & $247.2\pm 0.3$&$308.3\pm 0.5$&$202.8\pm 0.5$\\
{\it Total Light I=$0\, \oplus \,1$ ($1.875\to 3.68$)}&$\it 293.8\pm 1.1$&$\it 354.7\pm 1.2$&$\it 223.7\pm 0.7$\\
\rowcolor{greenli}{ Total Light I=$0\, \oplus \,1$ ($2m_\pi\to3.68$)}&$ 6786.1\pm 38.8$&$3038.4\pm 24.5$&$806.4\pm 4.6$\\
{\bf Charmonium} &&&\\
$J/\psi (1S)$ (NWA) &$65.1\pm 1.2$&$92.7\pm 1.8$&$73.5\pm 1.4$\\
$\psi (2S)$ (NWA) &$16.4\pm 0.6$&$26.0\pm 0.9$&$26.1\pm 0.8$\\
$\psi (3773)$ (NWA) &$1.7\pm 0.1$&$2.7\pm 0.2$&$2.9\pm 0.2$\\
{\it Total $J/\psi (NWA)$} &$\it 83.2\pm 1.4$&$\it 121.4\pm 2.0$&$\it 102.5\pm 1.6$\\
$3.69\to 3.86$ & $11.4\pm 1.0$&$18.3\pm 1.6$&$19.0\pm 1.6$\\
$3.86\to4.094$ & $16.6\pm 0.5$&$27.5\pm 0.8$&$30.9\pm 0.9$\\
$4.094\to4.18$ & $6.6\pm 0.2$&$11.2\pm 0.3$&$13.2\pm 0.4$\\
$4.18\to4.292$ & $6.5\pm 0.2$&$11.2\pm 0.4$&$13.7\pm 0.5$\\
$4.292\to4.54$ & $11.8\pm 0.6$&$20.7\pm 0.8$&$26.8\pm 1.1$\\
$4.54\to10.50$:  QCD $(u,d,s,c)$ & $92.0\pm 0.0$&$186.2\pm 0.0$&$458.7.3\pm 0.1$\\
{\it Total Charmonium ($3.69\to 10.50$)}&$\it 145.1\pm 1.3$&$\it 275.6\pm 2.0$&$\it 564.9\pm 2.2$\\
\rowcolor{greenli}{ Total Charmonium }&$ 228.1\pm 1.9$&$396.5\pm 2.8$&$664.8\pm 2.7$\\
 {\bf Bottomium} &&&\\
$\Upsilon (1S)$ (NWA) &$0.54\pm 0.02$&$1.25\pm 0.07$&$5.65\pm 0.29$\\
$\Upsilon(2S)$ (NWA) &$0.22\pm 0.02$&$0.51\pm 0.06$&$2.54\pm 0.29$\\
$\Upsilon (3S)$ (NWA) &$0.14\pm 0.02$&$0.33\pm 0.04$&$1.77\pm 0.23$\\
$\Upsilon(4S)$ (NWA) &$0.10\pm 0.01$&$0.23\pm 0.03$&$1.26\pm 0.16$\\
$\Upsilon(10.86\,\oplus\,11)$ (NWA) &$0.1\pm 0.06$&$0.20\pm 0.06$&$1.67\pm 0.39$\\
{\it Total Bottomium (NWA)}&$\it1.0\pm 0.1$&$\it 2.3\pm 0.1$&$\it 11.2\pm 0.5$ \\
$Z-pole $&-&-&29.2\,\cite{YND17}\\
$10.59\to2m_t$\,:  QCD $(u,d,s,c,b)$ & $22.4\pm 0.3$&$57.5\pm 0.1$&$1282.9\pm 1.2$\\
\rowcolor{greenli}{Total Bottomium }&$ 23.4\pm 0.3$&$59.8\pm 0.1$&$1323.3\pm1.3$\\
$2m_t\to\infty$\,: QCD $(u,d,s,c,b,t)$&0.03&0.08&-28.2\\
\hline
\rowcolor{yellow}\bf Total sum &\boldmath$7036.5\pm 38.9$&\boldmath$3494.8\pm 24.7$&\boldmath$2766.3\pm 4.5$\\

   \hline\hline
 
\end{tabular}}
 \caption{ 
  $a_\tau\vert^{hvp}_{l.o}$, $a_\mu\vert^{hvp}_{l.o}$ and $\alpha(M^2_z)$ within our parametrization of the compiled PDG\,\cite{PDG} $\oplus$ the recent CMD3\,\cite{CMD3}. }\label{tab:amu1} 
 \end{center}
\end{table*}
} 

\vspace*{-0.5cm}


\end{document}